\documentclass[11pt,reqno]{amsart}
\usepackage{amssymb, amsthm, amsmath,fancybox,graphicx,colordvi,t1enc}

\newtheorem*{namedtheorem}{\theoremname}
\newcommand{\theoremname}{testing}
\newenvironment{thm}[1]{\renewcommand{\theoremname}{#1}
\begin{namedtheorem}}{\end{namedtheorem}}
\newenvironment{rem}[1]{\renewcommand{\theoremname}{{\sc #1}}
\begin{namedtheorem}\em}{\end{namedtheorem}}
\newenvironment{lem}[1]{\renewcommand{\theoremname}{{\sc #1}}
\begin{namedtheorem}}{\end{namedtheorem}}

\def\uprho{\raise1pt\hbox{$\rho$}}
\def\sr{{\mathcal R}}
\def\R{{\mathbb R}}
\def\I{{\mathcal I}}
\def\sima{\mathrel{\overset{\rm\scriptscriptstyle
A}{\sim}}}
\def\simt{\mathrel{\overset{\rm\scriptscriptstyle
T}{\sim}}}
\def\mfr#1/#2{\hbox{${\frac{#1}{#2}}$}}
\begin{document}
    \markboth{\scriptsize{LY 09/03/02}}{\scriptsize{LY 09/03/02}}
\title{The Mathematical Structure of the Second Law of Thermodynamics}
\author{Elliott H.\ Lieb and Jakob Yngvason}
\thanks{E.H.L.'s work was partially supported by U.S.\ National
Science Foundation grants.
J.Y.'s work was partially
supported by the Adalsteinn Kristjansson Foundation,
University of Iceland.\\
\copyright  2002 by the authors.
Reproduction of this article, by any means, is permitted for
non-commercial purposes.}

\thanks{This article appears in {\it Current Developments in Mathematics, 2001},
International Press, Cambridge, 2002, pp. 89--130.}


\begin{abstract}
\noindent    The essence of the second law of classical thermodynamics
is the `entropy principle' which asserts the existence of an additive
and extensive  entropy function, $S$, that is defined for all
equilibrium states of thermodynamic systems
and whose increase characterizes the possible state changes under
adiabatic conditions. It is one of the few really fundamental physical
laws (in the sense that no deviation, however tiny, is permitted) and its
consequences are far reaching.  This principle is independent of 
models, statistical mechanical or otherwise, and can be understood
without recourse to Carnot cycles, ideal gases and other assumptions
about such things as `heat', `temperature', `reversible processes',
etc.,  as is usually done.  Also the well known formula of statistical 
mechanics, $S = -\sum p  \log p$, is not needed for the derivation of
the entropy principle.
  
This contribution is  partly a summary of our joint work (Physics
Reports, Vol.\ 310, 1--96 (1999)) where the existence and uniqueness
of $S$ is proved to be a consequence of certain basic properties of the
relation of adiabatic accessibility among equilibrium states.  We also
present some open problems and suggest directions for further study.

\end{abstract} \maketitle

\section*{Foreword}
At the conference ``Contemporary Developments in Mathematics'', hosted by
the MIT and Harvard University Mathematics Departments, November 16-17,
2001, one of us (E.H.L.) contributed a talk with the above title. It
was a review of our work \cite{[7]} on the mathematical foundations of
classical thermodynamics. An extensive summary of \cite{[7]} was published
in the AMS Notices \cite{[7A]}.  It was also published in \cite {GAFA}
and \cite{dresden} with  additional sections added in each.  A shorter
summary, addressed particularily to physicists, appeared in Physics
Today \cite{LiY3}. We include here an expanded version of the article
\cite{dresden}.  Section \ref{sec1} is primarily from \cite{[7A]} but is
augmented by proofs of all theorems.   The present version is  therefore
mathematically complete, but the original paper \cite{[7]} is recommended
for additional insights and extensive discussions.  Section  \ref{sec2}
is primarily from \cite {GAFA} and \cite{dresden}. Section \ref{sec3}
is mainly from \cite{LiY3}.

\section{A Guide to Entropy and the Second Law of
Thermodynamics} \label{sec1}

This article is intended for readers who, like us, were told that the
second law of thermodynamics is one of the major achievements of the
nineteenth century, that it is a logical, perfect and unbreakable law --
but who were unsatisfied with the `derivations' of the entropy principle
as found in textbooks and in popular writings.

A glance at the books will inform the reader that the law has `various
formulations' (which is a bit odd for something so fundamental) but
they all lead to the existence of an entropy function whose reason for
existence is to tell us which processes can occur and which cannot.
An interesting summary of these various points of view is in \cite{U}.
Contrary to convention, we shall refer to the existence of entropy as
{\em the} second law.  This, at least, is unambiguous.  The entropy
we are talking about is that defined by thermodynamics (and {\em not}
some analytic quantity, usually involving expressions such as $-p\ln p$,
that appears in information theory, probability theory and statistical
mechanical models).

Why, one might ask, should a mathematician be interested in the second
law of thermodynamics which, historically, had something to do with 
attempts to understand and improve the  efficiency of steam engines? The
answer, as we perceive it, is that the law is really an interesting
mathematical theorem about orderings on sets, with profound physical
implications. The axioms that constitute this ordering are somewhat
peculiar from the mathematical point of view and might not arise in the
ordinary  ruminations of abstract thought. They are special, but
important, and they are driven by considerations about the world, which
is what makes them so interesting.  Maybe an ingenious reader will find
an application of this same logical structure to another field of
science.

Classical thermodynamics, as it is usually presented, is based on three
laws (plus one more, due to Nernst, which is mainly used in low
temperature physics and is not immutable like the others). In brief,
these  are:

\begin{quote}
\noindent{\em The Zeroth Law,} which expresses the
transitivity of equilibrium, and which is often said to imply the
existence of temperature as a parametrization of equilibrium states. We
use it below but formulate it without mentioning temperature. In fact,
temperature makes no appearance here until almost the very end.

\noindent{\em The First Law,} which is conservation
of energy. It is a concept from mechanics  and provides the connection
between mechanics (and things like falling weights) and thermodynamics. 
We discuss this later on when we introduce simple   systems; the 
crucial usage of this law is that it allows   energy to be used
as one of the parameters describing the states of a simple system.

\noindent{\em The Second Law.} Three popular
formulations of this law are:

\begin{quote}
\noindent{\em Clausius\/:} No process is possible,
the sole result of which is that heat is transferred from a body to a
hotter one.

\noindent{\em Kelvin (and Planck)\/}: No process is
possible, the sole result of  which is that a body is cooled and work is
done.

\noindent{\em Carath\'eodory\/}: In any
neighborhood of any state there are states that cannot be reached from
it by an adiabatic process.
\end{quote}
\end{quote}

All three are supposed to lead to the entropy principle (defined below).
These steps can be found in many books and will not be trodden again
here. Let us note in passing, however, that the first two use concepts
such as hot, cold, heat, cool, that are intuitive but
have to be made precise before the statements are truly
meaningful. 
No one has seen
`heat', for example. The last (which uses the  term ``adiabatic
process\rq\rq, to be defined below) presupposes some kind of parametrization
of states by points in ${\mathbb R}^n$, and  the usual derivation of entropy
from it assumes some sort of differentiability; such assumptions are
beside the point as far as understanding the meaning of entropy goes.

The basic input in our analysis of the second law 
is a certain kind of ordering on a set
and denoted by 
$$ \prec 
$$ 
(pronounced `precedes').  It is transitive and reflexive as in A1, A2
below, but $X\prec Y$ and $Y\prec X$ does not imply $X=Y$,  so it is a
`preorder'.  The big  question is whether $\prec$ can be encoded in an
ordinary, real-valued  function on the set, denoted by $S$, such that if
$X$ and $Y$ are related by $\prec$, then $S(X)\leq S(Y)$ if and only if
$X\prec Y$.  The function $S$ is also required to be additive and
extensive in a sense that will soon be made precise.

A helpful analogy is the question: When can a vector-field, 
$\vec V(\vec x)$, on
${\mathbb R}^3$ be encoded in an ordinary function, $f(\vec x)$, whose gradient
is $\vec V$? The well-known answer is that a necessary and sufficient
condition is that ${\rm curl}\, \vec V=0$. Once $\vec V$ is observed to
have this property one thing becomes evident and important: It is
necessary to measure the integral of $\vec V$ only along some
curves -- not all curves -- in order to deduce the
integral along {\em all} curves.
The
encoding then has enormous predictive power about the nature of future
measurements of $\vec V$.  In the same way, knowledge of the function
$S$ has enormous predictive power in the hands of chemists, engineers
and others concerned with the ways of the  physical  world.

Our concern will be the existence and properties of $S$, starting from
certain natural axioms about the relation $\prec$.  We present our
results with slightly abridged versions of some
proofs, but full details, and a discussion of related
previous work on the foundations of classical thermodynamics, are given
in \cite{[7]}.
The literature on this subject is extensive and it is not
possible to give even a brief account of it here, except for mentioning
that the previous work closest to ours is that of
\cite{[6]}, and \cite{[2]}, (see also \cite{[4]},
\cite{[5]} and \cite{[9]}). (The situation is summarized more completely
in \cite{[7]}.)
These other approaches are also based on an
investigation of the relation $\prec$, but the overlap with our work is
only partial.  In fact, a major part of our work is the derivation of a
certain property (the ``comparison hypothesis\rq\rq\ below), which is taken as
an axiom in the other approaches.  It was a remarkable and largely
unsung achievement of Giles \cite{[6]} to realize the
full power of this property.

Let us begin the story with some basic concepts. 
\begin{itemize}
\item[1.] {\em Thermodynamic System\/}: Physically, this consists of
certain specified amounts of certain kinds of matter, e.g., a gram of
hydrogen in a container with a piston, or a gram of hydrogen and a gram
of oxygen in two separate containers, or a gram of hydrogen and two
grams of hydrogen in separate containers.  The system can be in various
states which, physically, are {\em equilibrium states}.  The space of
states of the system is usually denoted by a symbol such as $\Gamma$
and states in $\Gamma$ by $X,Y,Z,$ etc.
\end{itemize}

 Physical motivation aside, a state-space, mathematically, is
just  a set -- to begin with; later on we will be
interested in embedding state-spaces in some convex
subset of some $ {\mathbb R}^{n+1}$, i.e., we will introduce
coordinates.
As we said earlier, however, the
entropy principle is quite independent of  coordinatization, 
Carath\'eodory's principle notwithstanding.

\begin{itemize}
\item[2.] {\em Composition and scaling of states\/}:  The notion of
Cartesian product, $\Gamma_1 \times \Gamma_2$ 
corresponds simply to the two (or more) systems being side by side on
the laboratory table; mathematically it is just another system (called a
{\em compound system}), and  we regard  the state space  $\Gamma_1
\times \Gamma_2$ as the same  as $\Gamma_2 \times \Gamma_1$. 
Likewise, when forming multiple compositions of 
state spaces, the order and the grouping of the spaces is immaterial. 
Thus $(\Gamma_1 \times \Gamma_2)\times \Gamma_3$,  
$\Gamma_1 \times (\Gamma_2\times \Gamma_3)$ and 
$\Gamma_1 \times \Gamma_2\times \Gamma_3$ are to be identified as far 
as composition of state spaces is concerned.
Points in $\Gamma_1 \times \Gamma_2$ are denoted by pairs $(X,Y)$, 
and in $\Gamma_1 \times \cdots\times \Gamma_N$ by $N$-tuples 
$(X_{1},\dots, X_{N})$ as
usual. The subsystems comprising a compound system are physically
independent systems, but they are allowed to interact with each other
for a period of time and thereby alter each other's state.

 The concept of  scaling is crucial. It is this concept that
makes our thermodynamics inappropriate for microscopic objects like
atoms or cosmic objects like stars. 
For each state-space $\Gamma $ and number $\lambda>0$
there is another state-space, denoted by
$\Gamma^{(\lambda)}$ with points denoted by
\hbox{$\lambda X$}.
This space is
called a {\em scaled copy} of $\Gamma$.  Of course we identify
$\Gamma^{(1)} =\Gamma$ and $1X=X$.  We also require
$(\Gamma^{(\lambda)})^{(\mu)} = \Gamma^{(\lambda\mu)}$ and $\mu(\lambda
X) = (\mu\lambda)X$.  The physical interpretation of
$\Gamma^{(\lambda)}$ when $\Gamma$ is the space of one gram of hydrogen,
is simply the state-space of $\lambda$ grams of hydrogen. The state
$\lambda X$ is the state  of  $\lambda $ grams of hydrogen with the 
same `intensive' properties as $X$, e.g., pressure, while `extensive'
properties like energy, volume, etc., are scaled by a
factor $\lambda$ (by definition).  
\end{itemize}

 For any given $\Gamma$ we can form Cartesian product state  spaces
of the type $\Gamma^{(\lambda_1)} \times \Gamma^{(\lambda_2)} \times 
\cdots\times
\Gamma^{(\lambda_N)}$. These will be called {\em multiple scaled copies}
of $\Gamma$.  

The notation $\Gamma^{(\lambda)}$ should be regarded as merely a
mnemonic at this point, but later on, with the embedding of $\Gamma$
into $ {\mathbb R}^{n+1}$, it will literally be $\lambda\Gamma= \{\lambda
X:X\in \Gamma\}$ in the usual sense.  

\begin{itemize}
\item[3.] {\em 
Adiabatic accessibility\/}: Now we come to the ordering.
We say $X\prec Y$ (with $X$ and $Y$ {\em possibly 
in \underbar{\hbox{\strut different}} state-spaces}) if  $Y$ is {\em
adiabatically accessible} from $X$ according to the
definition below. Different state spaces can occur, e.g., if
there is mixing or a chemical reaction between two states of a compound
system to produce a state in a third system.
\end{itemize}

What does this  mean? Mathematically, we are just given a list of pairs
$X \prec Y$. There is nothing more to  be said, except that later on we
will assume that this list has certain properties that will lead to
interesting theorems about this list, and will lead, in turn,  to the
existence of an {\em entropy function}, $S$ characterizing the list.

The physical interpretation is quite another matter. In text books a
process taking $X$ to $Y$ is usually called adiabatic if it takes place
in `thermal isolation', which in turn means that `no heat is exchanged
with the surroundings'.  Such concepts (heat, thermal etc.) appear
insufficiently precise to us and we prefer the
following version, which is in the spirit of Planck's formulation of the
second law \cite{[8]} and avoids those concepts. 
Our definition of adiabatic accessibility might at first
sight appear to be less restrictive than the usual one,
but as discussed in \cite{[7]}, pp.\ 29 and 54, in the end anything
that we call an adiabatic process (meaning that $Y$ is
adiabatically accessible from $X$) can also be
accomplished in `thermal isolation' as the concept is
usually understood. 
Our definition has the great virtue
(as discovered by Planck) that it avoids having to
distinguish between work and heat -- or even having to
define the concept of heat.
We emphasize,
however, that the theorems do not require agreement with our physical
definition of adiabatic accessibility; other definitions are conceivably
possible. We emphasize also that we do not care about the temporal
development involved in the state change; we only care about the net
result for the system and the rest of the universe.

\begin{quote}
{\em A state $Y$ is adiabatically
accessible from a state $X$, in symbols $X\prec Y$, if it is possible to
change the state from $X$ to $Y$ by means of an interaction with some
device consisting  of some auxiliary system and a weight, in such a way
that the auxiliary system returns to its initial state at the end of the
process whereas the weight may have risen or fallen.}
\end{quote}

The role of the `weight' in this definition is merely to provide a 
particularly simple source (or sink) of mechanical energy. 
Note that an adiabatic process, physically, does not have to be gentle,
or `static' or anything of the kind. It can be arbitrarily violent and
destructive, so long as the system is brought back to equilibrium! 
The `device' need not be a well-defined mechanical contraption. 
It can be another thermodynamic system, and even a  gorilla jumping up
and down on the system, or a combination of these -- as
long as the device returns to its initial state. The device can have
intelligence, e.g., it can contain a clever scientist whose strategy
depends on the progress of the experiment. Only the initial state $X$
and the final state $Y$ matter.

An example might be useful here.  Take a pound of hydrogen in a
container with a piston. The states are describable by two numbers,
energy and volume, the latter being determined by the position of the
piston. Starting from some state, $X$, we can take our hand off the
piston and let the volume increase explosively to a larger one.  After
things have calmed  down, call the new equilibrium state $Y$.  Then
$X\prec Y$. Question: Is $Y\prec X$ true? Answer: No. To get from $Y$ to
$X$ adiabatically we would have to use some machinery and a weight,
with the machinery returning to its initial state, and there is no way
this can be done.  Using a weight we can, indeed, recompress the gas to
its original volume, but we will find that the energy is then larger
than its original value.

Let us write
$$
X\prec \prec Y \ \ \  {\rm if}\ \ \  X\prec Y \ \ \ {\rm but\ not}\ \ \
 Y\prec X \ {\rm (written\ }Y\not\prec X)\,. 
$$
In this case we say that we can go from $X$ to $Y$  by an
{\em irreversible adiabatic process}.
If $X\prec Y $ and $Y\prec X$ we say that $X$ and $Y$ are
{\em adiabatically equivalent } and write
$$
X\sima Y\,.$$

Equivalence classes under $\sima$ are called {\em adiabats}.
\begin{itemize}

\item[4.] {\em Comparability\/}: Given two states $X$ and $Y$ in two
(same or different)  state-spaces, we say that they are comparable if
$X\prec Y$ or $Y\prec X$ (or both). This turns out to be a crucial
notion. Two states are not always comparable; a necessary condition is
that they have the same material composition in terms of the chemical
elements. Example: Since water is ${\rm H}_{2} {\rm O}$ and  the atomic
weights of hydrogen and oxygen are 1 and 16 respectively,  the
states in the compound system of 2 gram of hydrogen and 16 grams of
oxygen are comparable with states in a system consisting of 18 grams of
water (but not with 11 grams of water or 18 grams of oxygen). 
\end{itemize}

Actually, the classification  of states into various state-spaces is
done  mainly for conceptual convenience.  The second law deals only
with states,  and the only thing we really have to know about any two
of them is whether or not they are comparable.  Given the relation
$\prec$ for all possible states of all possible systems, we can ask
whether this relation can be encoded in an entropy function according to
the following: 

\begin{thm}{Entropy principle} There is a real-valued function on all
states of all systems (including compound systems), called {\bf entropy}
and denoted by $S$ such that
\begin{enumerate}
\item[\rm a)] {\sc Monotonicity:} When $X$ and $Y$ are
comparable states then
\begin{equation}\label{1}
X\prec Y \hbox{ \ \ {\rm if and only if} \ \ }S(X)\leq S(Y)\,.
\end{equation}
\item[\rm b)] {\sc Additivity and extensivity:}
If $X$ and $Y$ are states of some $($possibly different$)$
systems and if $(X,Y)$ denotes the corresponding state in
the compound system, then the entropy is additive for
these states, i.e.,
\begin{equation}\label{2}
S(X,Y) = S(X) + S(Y) \,.  
\end{equation}
$S$ is also extensive, i.e., for or each $\lambda>0$ and
each state $X$ and its scaled copy
$\lambda X\in\Gamma^{(\lambda)}$, (defined in 2. above) 
\begin{equation}\label{3}
S(\lambda X)=\lambda S(X) \,.
\end{equation}
\end{enumerate}
\end{thm}

A  formulation logically equivalent to a),   not using 
the word `comparable', is the following pair of
statements:
\begin{eqnarray}\label{4}
X\sima Y\Longrightarrow S(X)=S(Y)\quad\rlap{\hbox{\rm
and}}\nonumber \\
X\prec\prec Y \Longrightarrow S(X) < S(Y)\,. 
\end{eqnarray}
The last line is especially noteworthy. It says that entropy must
increase in an irreversible  adiabatic  process.

The additivity of entropy in compound  systems is often just taken for
granted, but it is one of the startling conclusions of thermodynamics. 
First of all,  the content of additivity, (2), is considerably more far
reaching than one might think from the simplicity of the notation.
Consider four states $X,X',Y,Y'$ and suppose that $X\prec Y$ and
$X'\prec Y'$.  One of our axioms, A3, will be that then $(X,X')\prec
(Y,Y')$,  and (2) contains nothing new or exciting.  On the other
hand,  the compound system can well have an adiabatic process in which
$(X,X')\prec (Y,Y')$ but $X\not\prec Y$.  In this case, (2) conveys much
information. Indeed, by monotonicity, there will be many cases of this
kind because the inequality $S(X) + S(X') \leq S(Y) + S(Y')$ certainly
does not imply that $S(X) \leq S(Y)$. The fact that the inequality $S(X)
+ S(X') \leq S(Y) + S(Y')$ tells  us {\em exactly } which adiabatic
processes are allowed in the compound system (among  comparable states),
independent of any detailed knowledge of the manner in which the two
systems interact, is astonishing and is at the {\em heart of
thermodynamics.}  The second reason that (2) is startling is this: From
(1) alone, restricted to one system, the function $S$ can be replaced by
$29S$ and still do its job, i.e., satisfy (1). However, (2) says that it
is possible to calibrate  the entropies of all systems (i.e.,
simultaneously adjust all the undetermined multiplicative constants) so
that the entropy $S_{1,2}$ for a compound $\Gamma_1 \times \Gamma_2$ is
$S_{1,2}(X,Y) = S_1(X) + S_2(Y)$, even though systems 1 and 2 are
totally unrelated! 

We are now ready to ask some basic questions:
\begin{enumerate}
\item[Q1:]
Which properties of the relation $\prec$ ensure existence and
(essential) uniqueness of $S$?
\item[Q2:]
Can these properties be derived from simple physical premises?
\item[Q3:]
Which convexity and smoothness properties of $S$ follow from 
the premises?
\item[Q4:] Can temperature 
(and hence an ordering of states by ``hotness\rq\rq\ and ``coldness\rq\rq)
 be defined from $S$ and what are its properties?
\end{enumerate}

The answer to question Q1 can be given in the form of six axioms that
are reasonable,  simple, `obvious'  and unexceptionable.  An
additional, crucial assumption is also needed, but we call it a
`hypothesis' instead of an axiom because we show later how it can be
derived from some other axioms, thereby answering question Q2.  
\begin{enumerate}
\item[\bf A1.]  {\bf Reflexivity}.  $X \sima X$.
\item[\bf A2.]  {\bf Transitivity}. If $X \prec Y$ and $Y \prec Z$, 
then $X \prec Z$.
\item[\bf A3.] {\bf Consistency}. If $X \prec X^\prime$ and $Y
\prec Y^\prime$, then $(X,Y) \prec
(X^\prime, Y^\prime)$.
\item[\bf A4.] {\bf Scaling Invariance}. If $\lambda > 0$ and
$X \prec Y$,  then
$\lambda X \prec \lambda Y$.
\item[\bf A5.]  {\bf Splitting and Recombination}. 
$X \sima ((1-\lambda) X, \lambda X)$ for all $0 < \lambda <  1$. 
Note that the two state-spaces are different.
If $X\in\Gamma$, 
then the state space on the right side is 
$\Gamma^{(1-\lambda)} \times \Gamma^{(\lambda)}$.
\item[\bf A6.]  {\bf Stability}. If
$(X, \varepsilon Z_0) \prec (Y, \varepsilon Z_1)$
for some $Z_0,Z_1$ and a sequence of $\varepsilon$'s
tending to zero, then $X\prec Y$.
This axiom is a substitute for
continuity, which we cannot assume because there is no topology yet. It
says that  `a grain of dust cannot influence the set of adiabatic
processes'. 
\end{enumerate}

An important lemma is that (A1)--(A6) imply the {\em
cancellation law}, which is used in many proofs.
It says that for any three states $X,Y,Z$
\begin{equation}\label{cancellation}
 (X,Z)\prec (Y,Z) \Longrightarrow   X\prec Y
\end{equation}
\begin{proof}
We show that $(X,Z)\prec (Y,Z) $ implies  $(X,\mfr1/2Z)\prec (Y,\mfr1/2Z)$
and hence $(X,\mfr1/{2^n}Z)\prec (Y,\mfr1/{2^n}Z)$ for all $n=1,2,\dots$.
By the stability axiom, A6, this implies $X\prec Y$.

The argument for $(X,\mfr1/2Z)\prec (Y,\mfr1/2Z)$ is as follows:
\begin{eqnarray*}
  (X,\mfr1/2Z)&\sima &(\mfr1/2X,\mfr1/2X,\mfr1/2Z)\quad\text{(by A5
  and A3)}\\
    &\prec&(\mfr1/2X,\mfr1/2Y,\mfr1/2Z)\quad\text{(by $(X,Z)\prec
    (Y,Z)$, using  A3 and A4)}\\
    &\prec&(\mfr1/2Y,\mfr1/2Y,\mfr1/2Z)\quad\text{(again by $(X,Z)\prec
    (Y,Z)$, using  A3 and A4)}\\
    &\sima&(Y,\mfr1/2Z)\quad\text{(by A5 and A3).}
\end{eqnarray*}
 \end{proof}
    The next concept plays a key role in our treatment.
\begin{enumerate}
\item[\bf CH.]{\bf Definition:}
We say that the {\em Comparison
Hypothesis,} (CH), holds for a state-space  $\Gamma$  if  all
pairs  of states in   $\Gamma$ are comparable.
\end{enumerate}

Note that A3, A4 and A5 automatically extend comparability from a space
$\Gamma$ to certain other cases, e.g., $X\prec ((1-\lambda)Y,\lambda Z)$
for  all $0\leq \lambda \leq 1$  if  $X\prec Y$ and $X\prec Z$.  On
the other hand, comparability on $\Gamma$ alone does not allow us to
conclude that $X$ is comparable to $((1-\lambda)Y,\lambda Z)$ if $X\prec
Y$ but $Z\prec X$. For this, one needs CH on the product space
$\Gamma^{(1-\lambda)}\times \Gamma^{(\lambda)}$, which is not implied by
CH on $\Gamma$.

The significance of A1--A6 and CH is borne out by the
following theorem:

\begin{thm}{THEOREM 1 {\bf (Equivalence of entropy and A1-A6, given
CH)}}
The following are equivalent for a state-space $\Gamma$:
\begin{enumerate}
\item[\rm (i)]
 The relation $\prec$ between states in (possibly different)
multiple scaled copies of $\Gamma$
e.g.,
$\Gamma^{(\lambda_1)} \times \Gamma^{(\lambda_2)} \times \cdots\times
\Gamma^{(\lambda_N)}$,
is
characterized by an entropy function, $S$, on $\Gamma$ in the sense that
\begin{equation}\label{eq1-thm1}
(\lambda_1 X_1,\lambda_2 X_2,\dots)~\prec~(\lambda_1' X_1',\lambda_2'
X_2',\dots)
\end{equation}
is equivalent to  the condition that
\begin{equation}\label{eq2-thm1}\sum_i \lambda_i S(X_i)\leq
\sum_j\lambda_j'S(X_j')\end{equation}
whenever
\begin{equation}\label{eq3-thm1}\sum_i\lambda_i=\sum_j\lambda'_j\,.
\end{equation}
\item[\rm (ii)]
 The relation $\prec$ satisfies conditions A1--A6,
and CH holds for \underbar{every} multiple scaled copy
of $\Gamma$.
\end{enumerate}

This entropy function  on $\Gamma$ is unique up to
affine equivalence, i.e., $S(X)\rightarrow aS(X)+B$, with
$a>0$.
\end{thm}

\begin{proof}

The implication (i) $\Rightarrow$ (ii) is obvious.  To prove the
converse and also the uniqueness of entropy, pick two reference points
$X_0\prec\prec X_1$ in $\Gamma$.  (If there are no such points then
entropy is simply constant and there is nothing more to prove.) To
begin with, we focus attention on the
\lq strip\rq\ $\{X:\ X_{0}\prec X\prec  X_{1}\}$. (See Fig.\ 1.) 
In the
following it is important to keep in mind that, by axiom  A5,
$X\sima ((1-\lambda)X,\lambda X)$, so $X$ can be thought of as a point in
$\Gamma^{(1-\lambda)}\times \Gamma^{(\lambda)}$, for any $0\leq
\lambda\leq 1$.

Consider uniqueness first. If $S$ is any entropy function satisfying (i),
then necessarily
$S(X_0)<S(X_1)$, and $S(X)\in [S(X_0), S(X_1)]$. Hence there is a
unique $\lambda\in[0,1]$ such that \begin{equation}\label{four}
S(X)=(1-\lambda)S(X_{0})+\lambda S(X_{1}).  \end{equation} By (i), in
particular additivity and extensivity of $S$ and the fact that
$X \sima ((1-\lambda X, \lambda X) $, this is {\it equivalent} to
\begin{equation}\label{three} X\sima ((1-\lambda) X_{0},\lambda X_{1}).
\end{equation} Because \eqref{three} is a property of $X_0, X_1, X$ which
is independent of $S$, and because of the equivalence of \eqref{three} and
\eqref{four} for any entropy function, any other entropy function, $S'$
say,
must satisfy \eqref{four} with the {\it same} $\lambda$ but with $S(X_0)$
and $S(X_1)$ replaced by $S'(X_0)$ and $S'(X_1)$ respectively. This proves
that entropy is uniquely determined up to the choice of the entropy for
the two reference points. A change of this choice clearly amounts to an
affine transformation of the entropy function.

The equivalence of \eqref{three} and \eqref{four} provides also a
clue for constructing entropy: Using only the properties of the
relation $\prec$ one must produce a unique $\lambda$ satisfying
\eqref{three}. The uniqueness of such a $\lambda$, if it exists,
follows from the more general fact that
\begin{equation}\label{refpointS}((1-\lambda) X_{0},\lambda X_{1})
    \prec ((1-\lambda') X_{0},\lambda' X_{1})
    \end{equation}
is equivalent to
    \begin{equation}\label{lambdalambda}
        \lambda\leq \lambda'.
        \end{equation}
This equivalence follows from $X_0\prec\prec X_1$, using A4, A5 and the
cancellation law, \eqref{cancellation}.

To find $\lambda$  we consider
\begin{equation}\label{one}
\lambda_{\rm max}=\sup \{\lambda:\  ((1-\lambda) X_{0},\lambda
X_{1})\prec X\}
\end{equation}
and
\begin{equation}\label{two}
\lambda_{\rm min}=\inf \{\lambda:\ X\prec ((1-\lambda) X_{0},\lambda
X_{1})\}
\end{equation}
Making use of the stability axiom, A6, one readily shows that the sup
and inf are achieved, and hence
\begin{equation}\label{five}
((1-\lambda_{\rm max}) X_{0},
\lambda_{\rm max}
X_{1})\prec X
\end{equation}
and
\begin{equation}\label{six}
X\prec ((1-\lambda_{\rm min}) X_{0},
\lambda_{\rm min}
X_{1}).
\end{equation}
Hence, by A2,
\begin{equation}
((1-\lambda_{\rm max}) X_{0},
\lambda_{\rm max}
X_{1})\prec
((1-\lambda_{\rm min}) X_{0},
\lambda_{\rm min}
X_{1})
\end{equation}
and thus, (contrary to what the notation might suggest)
\begin{equation}
\lambda_{\rm max}\leq \lambda_{\rm min}.
\end{equation}
That
$\lambda_{\rm max}$ cannot be strictly smaller than $\lambda_{\rm
min}$ follows from the comparison hypothesis for the state spaces
$\Gamma^{(1-\lambda)}\times\Gamma^{(\lambda)}$: If 
$\lambda>\lambda_{\rm max}$, then $ ((1-\lambda) X_{0},\lambda X_{1})
\prec X$ can not hold,
and hence, by (CH) the alternative, i.e.,
\begin{equation}\label{eight} X\prec ((1-\lambda)
X_{0},\lambda X_{1})\end{equation}
must hold. Likewise, $\lambda<\lambda_{\rm
min}$ implies
\begin{equation}\label{nine}
  ((1-\lambda) X_{0},\lambda X_{1})\prec X.  
\end{equation} Hence, if $\lambda_{\rm max}<\lambda_{\rm min}$,  we
have produced a whole interval of $\lambda$'s satisfying
\eqref{three}. This contradicts the statement made earlier that
\eqref{three}
specifies at most one $\lambda$. At the same time we have shown that
$\lambda=\lambda_{\rm min}=\lambda_{\rm max}$ satisfies \eqref{three}.
Hence we can define the entropy by \eqref{four}, assigning some fixed,
but arbitrarily chosen values $S(X_0)<S(X_1)$ to the reference
points. For the special choice $S(X_0)=0$ and $S(X_1)=1$ we have the
{\bf basic formula for {\boldmath{$S$}}} (see Fig.\ 1):
\begin{equation} \label{ten}
\boxed{\   S(X)=\sup\big\{\lambda :((1-\lambda)X_0,\lambda X_1)
\prec X\big\}  \ , \   }
\end{equation}
or, equivalently,
\begin{equation}\label{tenx}
\boxed{ \  S(X)= \inf\big\{\lambda :X\prec((1-\lambda)X_0,\lambda
X_1)\big \} \  .  \    }
\end{equation}


The existence of $\lambda$ satisfying \eqref{four} may can be shown
 also for  $X$ outside the \lq strip\rq, i.e., for $X\prec X_{0}$
 or $X_{1}\prec X$, by simply interchanging the roles of $X,\ X_{0}$
 and $X_{1}$ in the considerations above.  For these cases we use the
convention
 that $(X, -Y)
\prec Z$ means $X \prec (Y, Z)$, and $(Y,0Z)=Y$. If $X\prec X_{0}$,
$\lambda$
in Eq.\ \eqref{four}
 will be $\leq 0$, and if $X_{1}\prec X$ it will be $\geq 1$.

  Our conclusion is that every $X\in\Gamma$ is equivalent, in the
  sense of $\sima$, to a scaled composition of the reference points
  $X_{0}$ and $X_{1}$.  By A5 this holds also for all points in
  multiply scaled copies of $\Gamma$, where by A4 we can assume that
  the total `mass' in \eqref{eq2-thm1} is equal to 1.  Moreover, by
  the definition of $S$, the left and right sides of \eqref{eq2-thm1}
  are just the corresponding compositions of $S(X_{0})$ and
  $S(X_{1})$.  To see that $S$ characterizes the relation on multiply
  scaled copies it is thus sufficient to show that 
  \eqref{refpointS} holds if and only if
\begin{equation}
(1-\lambda) S(X_{0})+\lambda S( X_{1})
    \leq (1-\lambda') S(X_{0})+\lambda'S(X_{1}).
    \end{equation}
Since    $S(X_{0})<S(X_{1})$ this is just the
equivalence of \eqref{refpointS} and
\eqref{lambdalambda} that was already mentioned.
 \end{proof}

\begin{figure}[htq]\begin{center}
      \includegraphics[width=9cm]{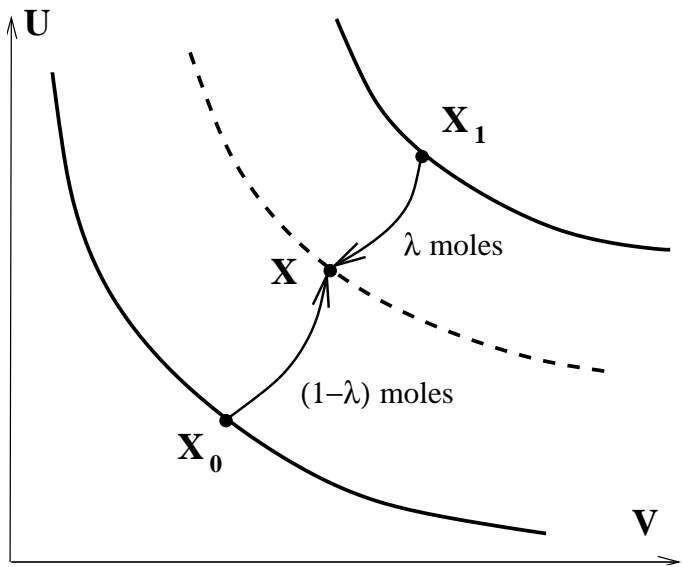}\end{center}
\caption{ The entropy of $X$ is, according to Eq.\ \eqref{ten}, 
  determined by the largest amount of $X_1$ that can be transformed
  adiabatically into $X$, with the help of a complementary amount of
  $X_0$. The coordinates U (energy) and V (work coordinates) are
  irrelevant for Eq.\ (21), but are important in the context of the
  {\em simple systems} to be discussed later.}
\label{fig1_cdm}
\end{figure}


\begin{rem}{Remarks} 1. The formula \eqref{ten} for entropy is
reminiscent of an old definition of heat by Laplace
and Lavoisier in terms of the amount of ice that a body
can melt: $S(X)$ is the maximal amount of substance in the state $X_{1}$ 
that 
can be transformed into the state $X$ with the help of a complimentary 
amount in the state $X_{0}$. According to \eqref{tenx} this is also 
the minimal amount of substance in the state $X_{1}$ that is needed to 
transfer a complementary amount in the state $X$ into the state $X_{0}$.
Note also that any $\lambda$ satisfying \eqref{eight} is an upper bound 
and any $\lambda$ satisfying \eqref{nine} is a lower bound to $S(X)$.

2. The construction of entropy in the proof above requires CH to
hold for the two-fold scaled products $\Gamma^{(1-\lambda)}\times
\Gamma^{\lambda}$. It is not sufficient that CH holds for $\Gamma$ alone,
but
in virtue of the other axioms it necessarily holds for all multiple
scaled products of $\Gamma$ if it holds for the two-fold scaled
products.

3.  Theorem 1 states the properties a binary relation on a set must
have in order to be characterized by a function satisfying our
additivity and extensivity requirements. The set is here the union of
all multiple scaled products of $\Gamma$.  In a quite different
context, this mathematical problem was discussed by I.N.\ Herstein and
J.\ Milnor already in 1953 \cite{HM} for what they call a \lq mixture
set\rq\ , which in our terminology corresponds to the union of all
two-fold scaled products $\Gamma^{(1-\lambda)}\times
\Gamma^{\lambda}$.  The main result of their paper is very similar to
Theorem 1 for this special case.
\end{rem}

Theorem 1 extends to products of multiple scaled
copies of different systems, i.e.\ to general  {\em compound}  systems.
This extension is an immediate consequence of the following theorem,
which is proved by applying \hbox{Theorem 1} to the
product of the system under consideration with some
standard reference system.

\begin{thm}{Theorem 2 {\bf(Consistent entropy scales)}} Assume that CH
holds for \underbar{all} compound systems.
For each system $\Gamma$ let
$S_{\Gamma}$ be some definite entropy function on $\Gamma$ in the sense
of Theorem 1.
Then there are constants $a_{\Gamma}$ and $B{(\Gamma)}$
such that the function $S$, defined for all states of all
systems by
\begin{equation}\label{19}
S(X)= a_{\Gamma} S_{\Gamma} (X)+ B{(\Gamma)}
\end{equation}
for $X\in \Gamma$, satisfies additivity $(2)$,
extensivity $(3)$, and monotonicity $(1)$ in the sense
that whenever $X$ and $Y$ are in the same state space
then
\begin{equation}\label{24}
X\prec Y \quad\quad \hbox{\rm if and only if} \quad\quad S(X)\leq
S(Y)\,.
\end{equation}

\end{thm}
\begin{proof}
The entropy function defined by \eqref{ten}
will in general not satisfy additivity and extensivity, because the
reference points can be quite unrelated for
the different state spaces $\Gamma$. Note that \eqref{ten} both fixes the
states
where the entropy is defined to be 0 (those that are $\sima X_{0}$) and
also an
arbitrary entropy unit for $\Gamma$ by assigning the value 1 to the
states $\sima X_{1}$. To obtain an additive and extensive entropy it
is first necessary to choose the points where the entropy is 0
in a way that
is compatible with these
requirements.

This can  be achieved by
considering the formal vector space spanned by all systems and choosing a
Hamel basis of systems $\{\Gamma_{\alpha}\}$ in this space such that every
system can be written uniquely as a scaled product of a finite number of
the $\Gamma_{\alpha}$'s. Pick an arbitrary  point in
each state space in the basis, and define for each
state space $\Gamma$ a corresponding point $X_{\Gamma} \in \Gamma$ as
a composition of these basis points. Then
\begin{equation}\label{additive}
    X_{\Gamma_{1}\times
    \Gamma_{2}}=(X_{\Gamma_{1}},X_{\Gamma_{2}}) \quad \text{and}\quad
    X_{t\Gamma}=t
    X_{\Gamma}.
    \end{equation}
Assigning the entropy 0 to these points $X_{\Gamma}$ is clearly
compatible with additivity and extensivity.

To ensure that the entropy unit is the same for all
state spaces, choose some fixed
space $\Gamma_{0}$ with  fixed reference points $Z_{0}\prec\prec
Z_{1}$. For any $\Gamma$ consider the product space
$\Gamma\times \Gamma_{0}$ and the entropy function
$S_{\Gamma\times \Gamma_{0}}$
defined by \eqref{ten} in this
space with reference points $(X_{\Gamma},Z_{0})$ and $(X_{\Gamma},Z_{1})$.
Then $X\mapsto S_{\Gamma\times \Gamma_{0}}(X,Z_{0})$
defines an entropy function on $\Gamma$ by
the cancellation law \eqref{cancellation}. It
it is additive and extensive by the properties
\eqref{additive} of
$X_{\Gamma}$,
and by Theorem 1 it is is related to any other entropy
function on $\Gamma$ by and affine transformation.

An {\bf explicit formula} for this additive and extensive entropy is
\begin{eqnarray}\label{addentr}
S(X)&=&\sup\{\lambda:\ (X_{\Gamma},\lambda Z_{1})\}
\prec (X,\lambda Z_{0})\\
&=&\inf\{\lambda:\ (X,\lambda Z_{0})\prec (X_{\Gamma},\lambda Z_{1})\},
\end{eqnarray}
because
\begin{equation}
(X,Z_0) \sima ( (1-\lambda ) (X_{\Gamma} , Z_0) , \lambda
(X_{\Gamma} , Z_1))
 \end{equation}
is equivalent to
\begin{equation}
(X,\lambda Z_0)\sima  (X_{\Gamma}, \lambda Z_1)).
 \end{equation}
by the cancellation law.
\end{proof}

Theorem 2 is what we need, except for the
question of mixing and chemical reactions, which is treated at the end
and which can be put aside at a first reading.  In other words, as
long as we do not consider adiabatic processes in which systems are
converted into each other (e.g., a compound system consisting of a
vessel of hydrogen and a vessel of oxygen is converted into a vessel
of water), the entropy principle has been verified.  If that is so,
what remains to be done, the reader may justifiably ask?  The answer
is twofold: First, Theorem 2 requires that CH holds for {\em all }
systems, including compound ones, and we are not content to  take
this as an axiom.  Second,
important notions of thermodynamics  such as
`thermal equilibrium'  (which will eventually
lead to a precise definition of `temperature' ) have not
appeared so far. We shall see that these two points (i.e., thermal
equilibrium and CH) are not unrelated.

As for CH, other authors, \cite{[6]}, \cite{[2]},
\cite{[4]} and \cite{[9]} essentially {\em postulate}
that it holds for all systems by making it axiomatic that
comparable states fall into equivalence classes.
(This
means that the conditions $X\prec Z$ and $Y\prec Z$ always imply that
$X$ and $Y$ are comparable: likewise, they must be comparable if $Z\prec
X$ and $Z\prec Y$.)
Replacing the concept of a `state-space' by that of an
equivalence class, the comparison hypothesis then holds in these other
approaches  {\em by  assumption}  for all state-spaces.   We, in
contrast, would like to derive CH from something that we consider more
basic.  Two ingredients will be needed: The analysis of certain special,
but commonplace systems called `simple systems' and some assumptions
about thermal contact (the `zeroth law') that will act as a kind of glue
holding the parts of a compound systems in harmony with each other.

A {\bf Simple System} is one whose state-space can be identified with
some open  convex subset of some ${\mathbb R}^{n+1}$ with a distinguished
coordinate denoted by $U$, called the {\em energy}, and additional
coordinates $V\in {\mathbb R}^{n}$, called  {\em work coordinates.}  The
energy coordinate is the way in which thermodynamics makes contact with
mechanics, where the concept of energy arises and is precisely defined.
The fact that the amount of energy in a state is independent of the
manner in which the state was arrived at is, in reality, the  first
law of thermodynamics.  A typical (and often the only) work coordinate
is the volume of a fluid or gas (controlled by a piston); other examples
are deformation coordinates of a solid or magnetization of a
paramagnetic substance.

Our goal is to show, with the addition of a few more axioms, that
CH holds for simple systems and their scaled products. In the process,
we will introduce more structure, which will capture the intuitive
notions of thermodynamics; thermal equilibrium is one.

Here, for the first time in our theory, coordinates are introduced. Up
to now state spaces were fairly abstract things; there was no topology.
Calculus, for example, played no role --- contrary to the usual 
presentation of classical thermodynamics. 
For simple systems we are talking about points in ${\mathbb R}^{n}$ and
we can thus talk about `open sets', `convexity', etc. In particular,
if we take a point $X$ and scale it to $tX$ then this scaling now has 
the usual concrete meaning it always has in ${\mathbb R}^{n}$, namely,
all coordinates of $X$ are multiplied by the positive number $t$. 
The notion of $+$, as in $X+Y$, had no meaning heretofore, but now it
has the usual one of addition of vectors in ${\mathbb R}^{n}$.

First, there is an axiom about convexity:
\begin{enumerate}
\item[\bf A7.] {\bf Convex combination.}
If $X$ and $Y$ are states of
a simple system and $t \in [0,1]$ then
\begin{equation}\big(t X, (1-t) Y\big) \prec t X +
(1-t)Y\,,
\end{equation}
in the sense of ordinary convex addition of points in ${\mathbb R}^{n+1}$.
A straightforward consequence of this axiom (and A5)  is that the
{\em forward  sectors}
\begin{equation}
A_{X}:=\{Y\in\Gamma :X\prec Y\}
\end{equation}
of states $X$ in a simple system $\Gamma$ are {\em convex}
sets. (See Fig.\ 2.)
\end{enumerate}


Another consequence is a connection between the existence of
irreversible processes and Carath\'eodory's principle
(\cite{[3]}, \cite{[1]}) mentioned above.

\begin{lem}{Lemma 1}  Assume (A1)--(A7) for
$\Gamma\subset {\mathbb R}^N$ and consider the following statements:
\begin{enumerate}
\item[\rm (a)] {\rm Existence of irreversible processes:} For every
$X\in\Gamma$
there is a $Y\in\Gamma$ with $X\prec\prec Y$.
\item[\rm
(b)] {\rm Carath\'eodory's principle:} In every neighborhood of every
$X\in\Gamma$
there is a $Z\in\Gamma$ with $X\not\prec Z$.
\end{enumerate}
Then {\rm (a)} $\Rightarrow$ {\rm (b)} always.
If the forward sectors in $\Gamma$ have interior points,
then {\rm (b)} $\Rightarrow$ {\rm (a)}.
\end{lem}
\begin{proof}
  Suppose that for some $X \in   \Gamma$ there is  a neighborhood,
${\mathcal
N}_X$ of $X$ such that ${\mathcal N}_X$ is contained in $A_X$.
(This is the negation of the statement that in every
neighbourhood of every $X$ there is a $Z$ such that $X\prec Z$ is
false.) Let $Y\in A_X$ be arbitrary.  By the convexity of $A_X$, which
is implied by axiom A7, $X$ is an interior point of a line segment
joining $Y$ and some point $Z\in {\mathcal N}_X$, and, again by A7,
\begin{equation}
(tZ,(1-t)Y) \prec X \sima (tX,(1-t)X)
\end{equation}
for some $t\in (0,1)$. But we also have that $(tX,
(1-t) X)\prec (tZ,(1-t) X) $ since $Z\in A_X$. This
implies, by the cancellation law and A4, that $Y\prec X$.
Thus we conclude that for some $X$, we have that $X\prec Y$ implies
$X\sima
Y$. This contradicts (a). In particular, we have shown that (a)
$\, \Rightarrow$(b).
\smallskip

Conversely, assuming that (a) is false, there is a point
$X_0$
whose forward sector is given by $A_{X_0} = \{ Y:Y\sima X_0 \}$. Let
$X$ be an interior point of $A_{X_0}$, i.e., there is a neighborhood
of $X$, ${\mathcal N}_{X}$, which is entirely contained in  $A_{X_0}$.
All points in ${\mathcal N}_{X}$ are adiabatically equivalent to $X_0$,
however,
and hence to $X$, since $X\in {\mathcal N}_{X}$. Thus, (b) is false.
\end{proof}

\begin{figure}[htq]\begin{center}
      \includegraphics[width=10cm]{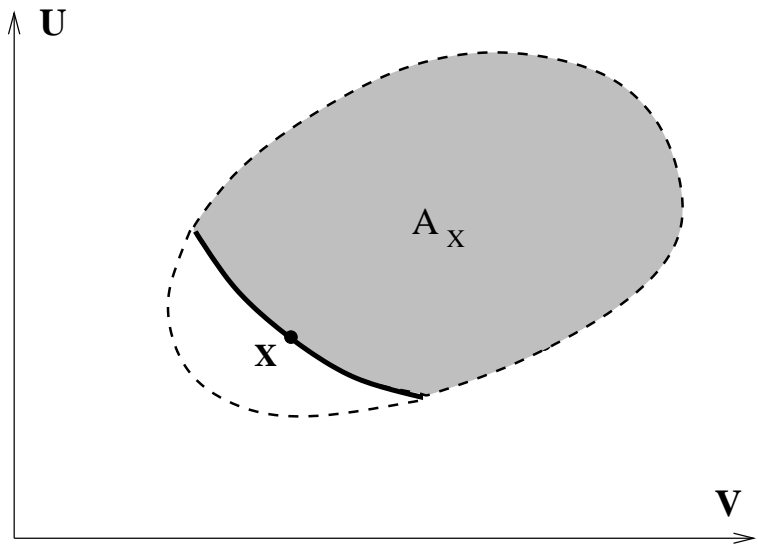}\end{center}
\caption{ The coordinates $U$ and $V$ of a simple system.
The state space (bounded by dashed line)  and the forward sector $A_X$
(shaded) of  a state  $X$ are convex, by axiom A7. The boundary of
$A_X$ (full line) is an adiabat, cf.\ Theorem 3.}
\label{figure2_cdm}
\end{figure}

We need three more axioms for simple systems, which will take
us into an analytic detour. The first of these
establishes (a) above.
\begin{enumerate}
\item[\bf A8.] {\bf Irreversibility.}
For each $X \in \Gamma$ there
is a point $Y \in \Gamma$ such that $X \prec\prec Y$. (This axiom
is implied by A14 below, but is stated here separately because
important conclusions can be drawn from it alone.)
\item[\bf A9.] {\bf Lipschitz tangent planes.}
For each $X\in \Gamma$
the {\em forward sector} $A_X=\{Y\in\Gamma:X\prec Y\}$ has a {\em
unique} support plane at $X$ (i.e., $ A_X$ has a {\em tangent plane} at
$X$).
The tangent plane is
assumed to be a {\em locally Lipschitz continuous} function of $X$,
in the sense explained below.
\item[\bf A10.] {\bf Connectedness of the boundary.}
The boundary
$\partial A_X$ (relative to the open set $\Gamma$) of every forward
sector $A_{X}\subset\Gamma$ is connected. (This is technical and
conceivably can be replaced by something else.)
\end{enumerate}

Axiom A8  plus
Lemma 1 asserts that every $X$ lies on the boundary $\partial A_X$ of
its forward sector. Although axiom A9 asserts that
the convex set, $A_X$, has a true tangent at $X$ only, it is an easy
consequence of axiom A2 that $A_X$ has a true tangent everywhere on its
boundary. To say that this tangent plane is locally Lipschitz
continuous means that if $X =
(U^0, V^0)$ then this plane is given by
\begin{equation}
U - U^0 + \sum^n_1 P_i (X) (V_i - V^0_i) = 0\,.
\end{equation}
with locally Lipschitz continuous functions $P_i$.
The function $P_i$ is called the generalized {\em pressure} conjugate to
the work coordinate $V_i$ .  (When $V_i$ is the volume, $P_i$ is the
ordinary pressure.)

Lipschitz continuity and connectedness is a well known
guarantee for uniqueness of the solution to the
coupled differential equations
\begin{equation}\label{33}
\tfrac{\partial u}{\partial V_j}(V)=-P_j\big(u(V),V\big)
\quad {\rm for} \ j=1,\dots ,n
\end{equation}
which describes the boundary
$\partial A_X$ of $A_X$.

With these axioms one can now prove that the comparison hypothesis
holds for the state space $\Gamma$ of a simple system:

\begin{thm}{Theorem 3 {\bf (CH for simple systems)}}  If $X$ and $Y$
are states of the same simple system, then either $X\prec Y$ or
$Y\prec X$. Moreover, $ X\sima Y\Longleftrightarrow Y\in\partial A_X\Longleftrightarrow X\in\partial A_Y$.

\end{thm}
\begin{proof} The proof is carried out in several
steps, which provide also further information about the forward
sectors.

{\it Step 1:  $A_{X}$ is closed.} We have to prove
that if $Y \in \Gamma$ is on the boundary of $A_X$ then $Y$  is in
$A_X$.  For this purpose we can assume that the set $A_X$ has full
dimension, i.e., the interior of $A_X$ is not empty.  If, on the
contrary, $A_X$ lay in some lower dimensional hyperplane then the
following proof would work, without any changes, simply by replacing
$\Gamma $ by the intersection of $\Gamma$ with this hyperplane.

Let $W$ be any point in the interior of
$A_X$. Since $A_X$ is convex, and $Y$ is on the boundary of $A_X$, the
half-open line segment joining $W$ to  $Y$ (call it $[W,Y)$, bearing in
mind that $Y\not\in [W,Y)$) lies in $A_X$.  The prolongation of this
line beyond $Y$
lies in the complement of $A_X$ and has at least one point (call it
$Z$) in $\Gamma$. (This follows from the fact that $\Gamma$ is open and
$Y\in \Gamma$.) For all sufficiently large integers $n$ the point $Y_n$
defined by
\begin{equation}
\mfr n/{n+1} Y_n +\mfr 1/{n+1} Z = Y
\end{equation}
belongs to $[W,Y)$.  We claim that $(X, \mfr1/n Z) \prec (Y, \mfr1/n Y)$.
If this is so then we are done because, by the stability axiom, A6,
$X\prec Y$.

To prove the last claim, first note that  $(X, \mfr1/n Z )\prec  (Y_n,
\mfr1/n Z)$ because $X\prec Y_n $ and by axiom A3. By scaling,
A4, the convex combination axiom A7, and (3.10)
\begin{equation}
\left(Y_n, \mfr1/n Z \right) \ = \mfr{n+1} /n \left(\mfr {n}/
{n+1} Y_n, \mfr1/{n+1} Z \right) \ \prec \ \mfr{n+1}/nY \ .
\end{equation}
But this last equals $ (Y, \frac{1}{ n} Y)$ by the splitting axiom, A5.
Hence $(X, \frac{1}{ n} Z) \prec (Y, \frac{1}{
n} Y)$.
\smallskip

{\it Step 2: $A_{X}$ has a nonempty interior.}  $A_{X}$ is a convex
set by axiom A7. Hence, if $A_{X}$ had an empty
interior it would necessarily be contained in a hyperplane.
[An illustrative  picture to keep in mind here
is  that $A_X$ is a closed, (two-dimensional) disc in ${\mathbb R}^3$ and
$X$ is some point inside this disc and not on its perimeter.
This disc is a closed subset
of ${\mathbb R}^3$ and
$X$ is on its boundary (when the disc is viewed as a subset of ${\mathbb
R}^3$).
The hyperplane is the plane in ${\mathbb R}^3$ that contains the disc.]

Any hyperplane containing $A_X$ is a support plane to $A_X$ at $X$, and
by axiom A9 the support plane, $\Pi_X$,  is unique, so $A_X\subset \Pi_X$.
If
$Y\in A_X$, then $A_Y\subset A_X\subset \Pi_X$ by transitivity, A2. By
the irreversibility axiom A8, there exists a $Y\in A_X$ such that
$A_Y\neq A_X$, which implies that the convex set $A_Y\subset \Pi_X$,
regarded as a subset of $\Pi_X$, has a boundary point in $\Pi_X$. If
$Z\in \Pi_X$ is such a boundary point of $A_Y$, then $Z\in A_Y$ because
$A_Y$ is closed. By transitivity, $A_Z\subset A_Y\subset \Pi_X$, and
$A_Z\neq \Pi_X$ because $A_Y\neq A_X$.

Now $A_Y$, considered as a subset of $\Pi_X$,  has an
$(n-1)$-dimensional supporting hyperplane at $Z$ (because $Z$ is a
boundary point).  Call this hyperplane $\Pi'_Z$. Since $A_Z\subset
A_Y$, $\Pi_Z'$ is a supporting hyperplane for $A_Z$,  regarded as a
subset of $\Pi_X$.  Any $n$-dimensional hyperplane in ${\mathbb R}^{n+1}$
that contains the $(n-1)$-dimensional hyperplane $\Pi'_Z\subset \Pi_X$
clearly supports $A_Z$ at $Z$, where $A_Z$ is now considered as a
convex subset of ${\mathbb R}^{n+1}$. Since there are infinitely many such
$n$-dimensional hyperplanes in ${\mathbb R}^{n+1}$, we have a
contradiction
to the uniqueness axiom A9.

\smallskip

{\it Step 3: $Y\in \partial A_X$ $\Rightarrow$ $X\in \partial A_Y$
and hence $A_{X}=A_{Y}$.}    We bring here only a sketch of the proof;
for details see \cite{[7]}, Theorems 3.5 and 3.6. First, using
the convexity axiom, A7, and the existence of a tangent plane of $A_X$
at $X$, one shows that the boundary points of $A_X$  can be written as
$(u_{X}(V),V)$, where $u_{X}$ is a solution to the equation system
\eqref{33}.  Here $V$ runs through the set
\begin{equation}
\rho_{X}=\{V:\ (U,V)\in\partial A_{X}\ \text{for some}\ U\}.
\end{equation}
Secondly, the solution of \eqref{33} that passes through $X$ is unique
by the Lipschitz condition for the pressure. In particular, if $Y\in
\partial
A_{X}$,  then $u_{X}$ must
coincide on $\rho_{Y}\subset \rho_{X}$ with the solution $u_{Y}$ through
$Y$. The proof is completed by showing that $\rho_{X}=\rho_{Y}$; this
uses that $\rho_{X}$ is connected by axiom A10 and also that $\rho_{X}$
is open. For the latter it is important that no tangent plane of
$A_{X}$ can be parallel to the $U$-axis, because of axiom A9.
\smallskip

{\it Step 4: $X\notin A_{Y}$ $\Rightarrow$ $Y\in A_{X}$.}
Let $Z$ be some point in the interior of $A_Y$ and consider  the line
segment $L$ joining $X$ to $Z$.
If we assume  $X\notin A_Y$ then part of $L$ lies outside
$A_Y$, and therefore $L$ intersects $\partial A_Y$ at some point $W\in
\partial A_Y$. By Step 3, $A_Y$ and $A_W$ are the same set, so
$W\prec Z$ (because $Y\prec Z$).
We claim that this implies $X\prec Z$ also.  This can be seen as
follows:

We have $W=tX+(1-t)Z$ for some $t\in (0,1)$.
By A7, A5, $W\prec Z$, and A3
\begin{equation}
(tX, (1-t)Z) \prec W \sima (tW,  (1-t) W)
\prec (tW,  (1-t)Z).
\end{equation}
By transitivity, A2, and the cancellation law, \eqref{cancellation},
$tX\prec
tW$. By scaling, A4, $X\prec W$ and hence, by A2, $X\prec Z$.

Since $Z$ was arbitrary, we learn that ${\rm Interior} (A_Y) \subset
A_X$.  Since $A_X$ and $A_Y$ are both closed by Step 1, this implies
$A_Y\subset A_X$ and hence, by A1, $Y\in A_X$.
\smallskip

{\it Step 5: $X\sima Y$ $\Longleftrightarrow$ $Y\in \partial A_{X}$.}
By Step 1, $A_{X}$ is closed, so $\partial A_{X}\subset A_X$. Hence,
if $Y\in \partial A_{X}$, then $X\prec Y$.  By Step 3, $Y\in \partial
A_X$ is equivalent to $X\in \partial A_Y$, so we can also conclude
that $Y\prec X$. The implication $\Longleftarrow$ is thus clear. On
the other hand, $X\sima Y$ implies $A_X=A_Y$ by Axiom A2 and thus
$\partial A_X=\partial A_Y$.  But $Y\in \partial A_{Y}$ by Axioms A1,
A8 and Lemma 1.  Thus the adiabats, i.e., the $\sima$ equivalence
classes, are exactly the boundaries of the forward sectors.
\end{proof}

%
\begin{figure}[htq]\begin{center}
      \includegraphics[width=8cm]{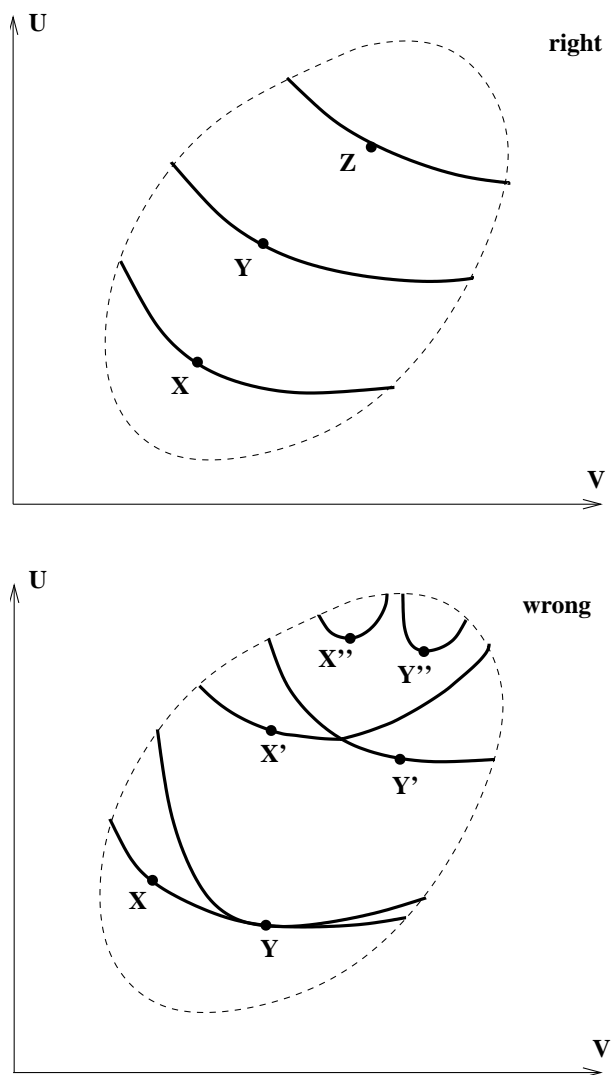}\end{center}
\caption{ This figure illustrates Theorem 3, i.e., the fact that 
forward sectors of a simple system are
nested.  The bottom figure shows what could, in principle, go 
wrong---but does not.}
\label{fiure3_cdm}
\end{figure}


\begin{rem}{Remark}  It can also be shown from our axioms that the
  orientation of forward sectors w.r.t.\ the energy axis is the same
  for {\em all} simple systems (cf.\ \cite{[7]}, Thms.\ 3.3 and 4.2).
  By convention we choose the direction of the energy axis so that the
  the energy always {\em increases} in adiabatic processes at fixed
  work coordinates.  When temperature is defined later, this will
  imply that temperature is always positive.  Since spin systems in
  magnetic fields are sometimes regarded as capable of having
  `negative temperatures' it is natural to ask what in our axioms
  excludes such situations.  The answer is: Convexity, A7, together
  with axiom A8.  The first would imply that if the energy can both
  increase and decrease in adiabatic processes, then also a state of
  maximal energy is in the state space.  But such a state would also
  have maximal entropy and thus violate A8.  From our point of view,
  `negative temperature' states should not be regarded as true
  equilibrium states.
\end{rem}

Before leaving the subject of simple systems let us remark on the
connection with Carath\'eo\-dory's development. The point of  contact is
the fact that $X\in \partial A_X$. We assume that $A_X$ is convex and
use transitivity and Lipschitz continuity to arrive, eventually, at
Theorem 3.
Carath\'eo\-dory uses Frobenius's theorem, plus assumptions about
differentiability to conclude the existence -- locally
-- of a surface containing $X$.
Important {\em global} information, such as Theorem 3,
are then not easy to obtain without further assumptions,  as discussed,
e.g., in \cite{[1]}.

The next topic is {\em thermal contact}  and the zeroth law,  which
entails the very special assumptions about $\prec$ that we mentioned 
earlier.  It will enable us to establish CH for products of several
systems, and thereby show, via Theorem 2, that entropy exists and is
additive. Although we have established CH for a simple system, $\Gamma$,
we have not yet established CH even for a product of two copies of
$\Gamma$. This is needed in the definition of $S$ given in (9).  The
$S$ in (9)  is determined up to an affine shift and we want to be able
to calibrate the entropies (i.e., adjust the multiplicative and additive
constants) of all systems so that they work together to form a global
$S$ satisfying the entropy principle.  We need five more axioms.  They
might look a bit abstract, so a few words of introduction might be
helpful.

In order to relate systems to each other, in the hope of establishing CH
for compounds, and thereby an additive entropy function, some way must
be found to put them into contact with each other. Heuristically, we
imagine two simple systems (the same or different) side by side, and fix
the work coordinates (e.g., the volume) of each. Bring them into \lq 
thermal contact\rq\ (e.g., by linking them to each other 
with a copper thread)  and wait for equilibrium to be established. The total
energy $U$ will not change but the individual energies, $U_1$ and $U_2$
will adjust to  values that depend on $U$ and the work coordinates. 
This new system (with the thread permanently connected) then behaves
like a simple system (with one energy coordinate) but with several work
coordinates (the union of the two work coordinates). 
Thus, if we start initially with $X=(U_1, V_1)$ for
\hbox{system 1} and $Y=(U_2, V_2)$  for system 2, and
if we end up with $Z=(U, V_1, V_2)$ for the new system, we
can say that $(X, Y) \prec Z$.
This holds for every choice of  $U_1
$ and $U_2$ whose sum is $U$. Moreover, after thermal equilibrium is
reached, the two systems can be disconnected, if we wish, and once more
form a compound system, whose component parts we say are in thermal
equilibrium. That this is transitive is the zeroth law.  

Thus, we cannot only make compound systems consisting of
independent subsystems (which can interact, but separate
again), we can also make a new simple system out of two
simple systems. 
To do this an energy
coordinate has to disappear, and thermal contact does this for us. This
is formalized in the following two axioms.  
\begin{enumerate}
\item[\bf A11.] {\bf Thermal join.}
For any two simple systems with state-spaces $\Gamma_1$
and $\Gamma_2$, there is another {\em simple} system,
called the {\em thermal join} of $\Gamma_1$ and
$\Gamma_2$, with state-space
\begin{equation}
\Delta_{12}=\big\{(U,V_1,V_2):U=U_1+U_2\;{\rm with}\; 
(U_1,V_1)\in \Gamma_1\,,\ (U_2,V_2)\in\Gamma_2\big\}\,.
\end{equation}
If $X=(U_1,V_1)\in \Gamma_{1}$, and $Y=(U_2,V_2)\in\Gamma_{2}
$ we define
\begin{equation}
 \theta(X,Y):=(U_1+U_2,V_1,V_2) \in \Delta_{12}.  
\end{equation}    
It is assumed that the formation of a thermal join is an adiabatic 
operation for the compound system, i.e.,
\begin{equation}
(X,Y)\prec \theta(X,Y).
\end{equation}
\item[\bf A12.] {\bf Thermal splitting.}
For any point $Z\in\Delta_{12}$ there is at
least one pair of states, $X\in\Gamma_1$,
$Y\in\Gamma_2$, such that
\begin{equation}
Z=\theta(X,Y)\sima (X,Y)
\end{equation}
\end{enumerate}

\noindent{\bf Definition.} If $\theta(X,Y)\sima (X,Y)$ we say that 
the states $X$ and $Y$ are in {\em
thermal equilibrium} and write 
\begin{equation}
X\simt Y\,.
\end{equation}

A11 and A12 together say that for each choice of the individual work
coordinates there is a way to divide up the energy $U$ between the two
systems in a stable manner.  A12 is the stability statement, for it
says that joining is reversible, i.e., once the equilibrium has been
established, one can cut the copper thread and retrieve the two
systems back again, but with a special partition of the energies. 
This reversibility allows us to think of the thermal join, which is a
simple system in its own right, as a special subset of the product
system, $\Gamma_1 \times \Gamma_2$, which we call the {\em thermal
diagonal}.

Axioms A11 and A12, together with the general axioms A4, A5, A7 and
our assumption that a compound state
$(X,Y)$ is identical to $(Y,X)$, imply that the relation $\simt$ is
refelxive and symmetric:

 \begin{lem}{lemma 2 }
 
\begin{enumerate}
    
\item[\rm (i)]  $X\simt X$.
\item[\rm (ii)] If $X\simt Y$, then $Y\simt X$. 
\end{enumerate}
\end{lem}

\begin{proof} (i) Let $X=(U,V)$. Then, by A11,
\begin{equation}(X,X)\prec \theta(X,X)=(2U,V,V).
 \end{equation}
 By A12, this is, for some $U'$ and $U^{\prime\prime}$ with 
$U'+U^{\prime\prime}=2U$,
 \begin{equation}
    \sima 
    ((U',V),(U^{\prime\prime},V)), 
  \end{equation}
  which,  using A4, A7, and finally A5,  is
  \begin{equation}
    \sima 
    2(\mfr1/2(U',V), \mfr1/2(U^{\prime\prime},V))
    \prec 2(\mfr1/2(U'+U^{\prime\prime}),V)=2X\sima (X,X). 
    \end{equation}
     Hence 
    $(X,X)\sima \theta(X,X)$, i.e., $X\simt X$.

(ii)  By A11 and A12 we have quite generally
\begin{equation}\theta(X,Y)
=(U_X+U_Y,V_X,V_Y)\sima ((U',V_X),
(U^{\prime\prime},V_Y))\end{equation}
for some $U'$, $U^{\prime\prime}$ with 
$U'+U^{\prime\prime}=U_X+U_Y=U_Y+U_X$. Since composition of states is 
commutative (i.e., $(A,B)=(B,A)$, as we stated when explaining the basic
concepts)  we obtain, using
A11 again,
\begin{equation} ((U',V_X),
(U^{\prime\prime},V_Y))=(
(U^{\prime\prime},V_Y),(U',V_X))\prec (U^{\prime\prime}+U',V_Y,V_X)=\theta(Y,X).
\end{equation}
Interchanging $X$ and $Y$ we thus have $\theta(X,Y)\sima 
\theta(Y,X)$. Hence, if 
$X\simt Y$, i.e., $(X,Y)\sima\theta(X,Y)$, then  $(Y,X)=(X,Y)\sima 
\theta(X,Y)\sima\theta(Y,X)$, i.e., $Y\simt X$.
\end{proof}

\begin{rem} {Remark} Instead of  presenting a formal proof 
of the symmetry of $\simt$, it
might seem more natural to simply identify $\theta(X,Y)$ with
$\theta(Y,X)$ by an additional axiom.  This could be justified both
from the physical interpretation of the thermal join, which is
symmetric in the states (connect the systems with a copper thread),
and also because both joins have the same energy and the same work
coordinates, only written in different order.  But since we really
only need that $\theta(X,Y)\sima \theta(Y,X)$ and this follows from
the axioms as they stand, it is not necessary to postulate such an
identification.  Likewise, it is not necessary to
identify $\theta(\theta(X,Y),Z)$ with $\theta(X,\theta(Y,
Z))$ formally by an axiom, because we do not need it. It is possible,
using the present axioms,  to
prove $\theta(\theta(X,Y),Z) \sima \theta(X,\theta(Y,
Z))$, but we do not do so since we do not need this either. 
\end{rem}

We now come to the famous zeroth law, which says that the thermal
equilibrium is transitive, and hence (by Lemma 2) an equivalence
relation.

\begin{enumerate}
\item[\bf A13.] {\bf Zeroth law of thermodynamics.} If $X\simt Z$ and 
$Z\simt Y$ then $X\simt Y$.  
\end{enumerate}

The zeroth law is often
taken to mean that the equivalence classes can be labeled by an
`empirical' temperature, but we do not want to mention temperature at
all at this point.  It will appear later.

There are two more  axioms about thermal contact, but before we state 
them we draw two simple conclusions from A11,  A12 and A13. 

\begin{lem}{lemma 3}
\begin{enumerate}
\item[\rm (i)]  If $X\simt Y$, then $\lambda X\simt \mu Y$ for all 
$\lambda,\mu>0$.
\item[\rm (ii)] If $X\simt Y$ and $Z\simt X$, then $Z\simt\theta(X,Y)$.
\end{enumerate}
\end{lem}
\begin{proof} 
    (i) By the zeroth law, A13, it suffices to show that $\lambda 
    X\simt X$ and $\mu Y\simt Y$. For this we use similar
    arguments as in Lemma 2(i):
    \begin{equation}(\lambda X,X)\prec \theta(\lambda 
    X,X)=((1+\lambda)U,\lambda V,V)\sima 
    ((\lambda U',\lambda V),(U^{\prime\prime},V))
        \end{equation}
    with $\lambda U'+U^{\prime\prime}=(1+\lambda)U$. By A4, A7, and
    A5 this is
    \begin{equation}
\sima 
    (1+\lambda)(\mfr\lambda/{1+\lambda}(U',V), \mfr1/{1+\lambda}
    (U^{\prime\prime},V))
    \prec (1+\lambda)(\mfr1/{1+\lambda}(\lambda U'+U^{\prime\prime}),V)=
    (1+\lambda)X\sima (\lambda X,X).    
    \end{equation}
    
    (ii) By the zeroth law, it suffices to show that 
    $X\simt\theta(X,Y)$,  i.e.,
    \begin{equation}\label{ass}(X,\theta(X,Y))\sima \theta (X,\theta(X,Y)).
\end{equation}
    The  left side of this equation is $\prec$ the right side by A11, so we 
    need only show
    \begin{equation}\label{ass2}
    \theta(X,\theta(X,Y))\prec
        (X,\theta(X,Y)).
\end{equation}
    Now, since $X\simt Y$ and hence also $2X\simt Y$ by (i), the right side 
    of \eqref{ass2} is 
\begin{equation}\label{ass1}
   (X,\theta(X,Y)) \sima (X,(X,Y))=((X,X),Y)\sima (2X,Y)\sima \theta 
(2X,Y)=(2U_{X}+U_{Y},2V_{X},V_{Y}). 
\end{equation} 
    (Here A5, A4 and A3 have been used, besides (i)). On the other hand, 
    using  A12 twice as well as A3, we have for some 
    $X'=(U',V_{X})$, 
    $X^{\prime\prime}=(U^{\prime\prime},V_{X})$ and 
    $Y'=(U^{\prime\prime\prime},V_{Y})$ with 
    $U'+U^{\prime\prime}+U^{\prime\prime\prime}=2U_{X}+U_{Y}$:
\begin{equation}\label{}
    \theta(X,\theta(X,Y))\sima (X',(X^{\prime\prime},Y'))=
    ((X',X^{\prime\prime}),Y') \  .
\end{equation}
By convexity A7 and scaling A4, as above, this is
\begin{equation}
\prec 
    (X'+X^{\prime\prime},Y')\prec
    \theta(X'+X^{\prime\prime},Y')=(2U_{X}+U_{Y},2V_{X},V_{Y}).
\end{equation}
But this is $\sima (X,\theta(X,Y))$ by \eqref{ass1}.
\end{proof}

We now turn to the remaining two axioms about thermal contact.

A14 requires that for every adiabat (i.e., an equivalence class
w.r.t.\ $\sima$) there exists at least one isotherm (i.e., an
equivalence class w.r.t.\ $\simt$), containing points on both sides of
the adiabat.  Note that, for each given $X$, only two points in the
entire state space $\Gamma$ are required to have the stated property. 
This assumption essentially prevents a state-space from breaking up
into two pieces that do not communicate with each other.  Without it,
counterexamples to CH for compound systems can be constructed, cf.\ \cite{[7]}, Section 4.3.  A14
implies A8, but we listed A8 separately in order not to confuse the
discussion of simple systems with thermal equilibrium.


A15 is a technical and perhaps can be eliminated. Its physical
motivation is that a sufficiently large copy of a system can act as a
heat bath for other systems. When temperature is introduced later, A15
will have the meaning that all systems have the same temperature range. 
This postulate is needed if we want to be able to bring every system
into thermal equilibrium with every other system.
\begin{enumerate}
\item[\bf A14.] {\bf Transversality.}
If
$\Gamma$ is the state space of a simple system and if $X \in \Gamma$,
then there exist states $X_0\simt X_1$ with $X_0\prec\prec X\prec\prec
X_1$.
\item[\bf A15.] {\bf Universal temperature range.}
If $\Gamma_1$ and
$\Gamma_2$ are state spaces of simple systems then,  for every
$X\in\Gamma_1$ and every $V$ belonging to the projection of
$\Gamma_2$ onto the space of its work coordinates,  there is a
$Y\in\Gamma_2$ with work coordinates $V$ such that $X\simt Y$.
\end{enumerate}

\begin{figure}[htq]\begin{center}
      \includegraphics[width=10cm]{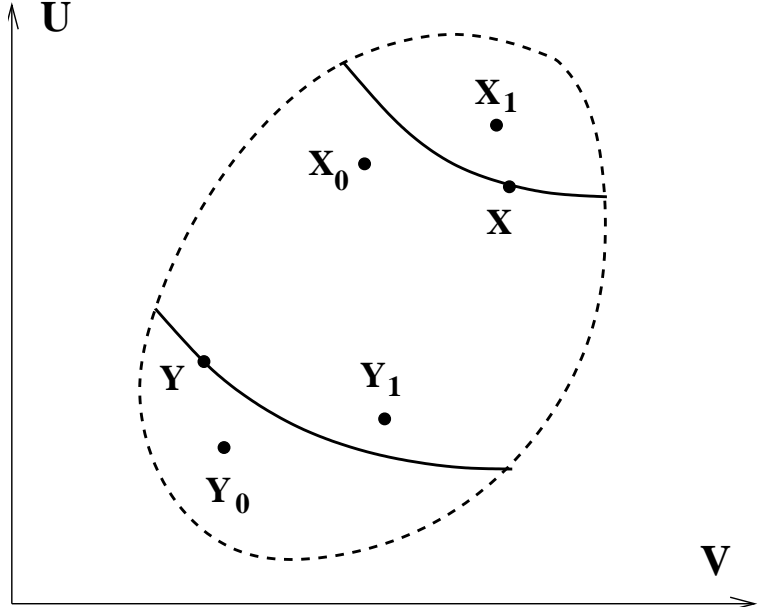}\end{center}
\caption{Transversality, A14, requires that on
on each side of the adiabat through any point $X$ there are
points, $X_0$ and $X_1$, that are in thermal equilibrium with each other.}
\label{figure4_cdm}
\end{figure}


The reader should note that the concept `thermal contact' has
appeared, but not temperature or hot and cold or anything resembling
the Clausius or Kelvin-Planck formulations of the second law. 
Nevertheless, we come to the main achievement of our approach: {\em
With these axioms we can establish CH for products of simple systems}
(each of which satisfies CH, as we already know).  The proof has two
parts.  In the first, we consider multiple scaled copies of the {\it
same} simple system and use the thermal join and in particular
transversality to reduce the problem to comparability within a single
simple system, which is already known to hold by Theorem 3.  The basic
idea here is that with $X, X_{0}, X_{1}$ as in A14, the states
$((1-\lambda)X_{0}, \lambda X_{1})$ and $((1-\lambda)X, \lambda X)$
can be regarded as states of the {\em same} simple system and are,
therefore, comparable.  {\em This is the key point needed for the
construction of $S$, according to $(9)$}.  The importance of
transversality is thus brought into focus.  In the second part we
consider products of {\it different} simple systems.  This case is
more complicated and requires all the axioms A1--A14, in particular
the zeroth law, A13.

\begin{lem}{lemma 4 {\rm (CH in multiple scaled copies of a simple system)}}
For any simple system $\Gamma$, all states of the form 
$(\lambda_{1}Y_{1}, 
\lambda_{2}Y_{2}\dots)$ with $Y_{i}\in\Gamma$ and 
$\sum_{i}\lambda_{i}$ fixed  are comparable.
    \end{lem}
\begin{proof} By scaling invariance of the relation $\prec$ (Axiom A4) we may 
assume that $\sum_{i}\lambda_{i}=1$. Now suppose 
$Y_{1},\dots,Y_{N}, Y_{1}',\dots,Y_{M}'\in \Gamma$, and 
$\lambda_{1},\dots,\lambda_{N}, 
\lambda_{1}',\dots,\lambda_{M}'\in{\mathbb R}$ with $\sum_{i}\lambda_{i}=
\sum_{j}\lambda_{j}'=1$. We shall show that for some 
$\bar X_{0},\bar X_{1}\in\Gamma$ with $\bar X_{0}\prec\prec 
\bar X_{1}$ and some $\lambda,\lambda'\in {\mathbb R}$
\begin{eqnarray}
  (\lambda_{1}Y_{1},\dots,\lambda_{N}Y_{N})&\sima& 
  ((1-\lambda)\bar X_{0},\lambda \bar X_{1}),\label{44}\\ 
  (\lambda_{1}'Y_{1}',\dots,\lambda_{M}'Y_{M}')&\sima& 
  ((1-\lambda')\bar X_{0},\lambda' \bar X_{1}).\label{45}
\end{eqnarray}
 This will prove the lemma, since we already know from the 
 equivalence of \eqref{refpointS} and \eqref{lambdalambda} that the right sides of \eqref{44} 
 and \eqref{45} are comparable.

 It was already noted that if $X_{0}\prec\prec X_{1}$ and $X_{0}\simt
 X_{1}$, then every $X$ in the \lq strip\rq\ $\Sigma (X_0, X_1):=\{X:\
 X_{0}\prec X\prec X_{1}\}$ is comparable to
 $((1-\lambda)X_{0},\lambda X_{1})$ for any $0\leq \lambda\leq 1$, due
 to the axioms A5, A11, A12, and Theorem 3.  This implies in the same
 way as in the proof of Theorem 1 that $X$ is in fact {\em
 adiabatically equivalent}  to $((1-\lambda)
 X_{0},\lambda X_{1})$ for some $\lambda$.  (Namely,
 $\lambda=\lambda_{\rm max}$, defined by \eqref{one}.)  Moreover, if
 each of the points $Y_{1}, Y_{2}\dots,Y'_{M-1}, Y'_{M}$ is
 adiabatically equivalent to such a combination of a {\it common} pair
 of points $\bar X_{0}\prec\prec \bar X_{1}$ (which need not be in
 thermal equilibrium), then Eqs.\ \eqref{44} and \eqref{45} follow
 easily from the recombination axiom A5.  The existence of such a
 common pair of reference points is proved by the following stepwise
 extension of `local' strips defined by points in thermal equilibrium.

By the transversality property, A4, the whole state space $\Gamma$ can
be covered by strips $\sum (X^{(i)}_0, X^{(i)}_1)$ with $X^{(i)}_0
\prec\prec X^{(i)}_0$ and $X^{(i)}_0 \simt X^{(i)}_1$.  Here $i$
belongs to some index set.  Since all adiabats $\partial A_X$ with $X
\in \Gamma$ are relatively closed in $\Gamma$ we can even cover each
$X$ (and hence $\Gamma$) with the {\it open} strips
$\mathop{{\sum}_i}\limits^o := \sum\limits^o (X^{(i)}_0, X^{(i)}_1) =
\{ X: X^{(i)}_0 \prec\prec X \prec\prec X^{(i)}_0 \}$.  Moreover, any
compact subset, $C$, of $\Gamma$ is covered by a finite number of such
strips $\mathop{{\sum}_i}\limits^o, i = 1, \dots , K$, and if $C$ is
connected we may assume that $\mathop{{\sum}_i}\limits^o \cap
\mathop{{\sum}_{i+1}}\limits^{o{\phantom{111}}} \not= \emptyset$.  In
particular, this holds if $C$ is some polygonal path connecting the
points $Y_{1},\dots,Y_{N}, Y_{1}',\dots,Y_{M}'$.

By Theorem 3, the points $X^{(1)}_0, X^{(1)}_1\dots X^{(K)}_0, X^{(K)}_1$, 
can be ordered according to the 
relation $\prec$, and there is no restriction to assume that
\begin{equation}
 X^{(i)}_0\prec\prec X^{(i+1)}_0\prec\prec X^{(i)}_1\prec\prec X^{(i+1)}_1.
\end{equation}  
Let $\bar X_{0}=X^{(1)}_0$ denote the \lq smallest\rq\ and 
$\bar X_{1}=X^{(K)}_1$  the \lq largest\rq\ of these points. We claim that 
every one of the points 
$Y_{1},\dots,Y_{N}, Y_{1}',\dots,Y_{M}'$ is adiabatically equivalent 
to a combination of $\bar X_{0}$ and $\bar X_{1}$. This is based on 
the following general fact: 

Suppose $X_0 \prec\prec
X_1, \,X'_0 \prec\prec X^\prime_1$ and 
\begin{equation}\label{46} X_1 \sima ((1 - \lambda_1)
X^\prime_0, \lambda_1 X^\prime_1), \quad \quad 
X^\prime_0 \sima ((1-
\lambda_0) X_0, \lambda_0 X_1). \end{equation} 
If
\begin{equation}\label{46a}X \sima ((1 -
\lambda) X_0, \lambda X_1) ,\end{equation} 
then 
\begin{equation}\label{47} X \sima ((1 -
\mu) X_0, \mu X^\prime_1) 
\end{equation}
with \begin{equation}\mu = \frac{\lambda \lambda_1}
{1 - \lambda_0 + \lambda_0 \lambda_1},
\end{equation} and if
\begin{equation}X \sima ((1 -
\lambda') X'_0, \lambda' X'_1) ,\end{equation} 
then 
\begin{equation}\label{48} X \sima ((1 -
\mu') X_0, \mu' X'_1) 
\end{equation}
with \begin{equation}\mu' = \frac{\lambda' (1-\lambda_{0})+
\lambda_{0}
\lambda_1
}{1 - \lambda_0 + \lambda_0 \lambda_1}.
\end{equation}
The proof of \eqref{47} and \eqref{48} is simple arithmetics, using the 
splitting and 
recombination axiom, A5, and the cancellation law, Eq.\ 
\eqref{cancellation}. Applying this successively for $i=1,\dots,K$ with
$X_{0}=X^{(1)}_0$, $X_{1}=X^{(i)}_1$,  $X'_{0}=X^{(i+1)}_0$, 
$X'_{1}=X^{(i+1)}_1$, proves that any 
$X\in\Sigma(X^{(1)}_0,X^{(K)}_1)$ is adiabatically equivalent
to a combination of $\bar X_{0}=X^{(1)}_0$ and $\bar 
X_{1}=X^{(K)}_1$. As already noted, this is precisely what is neded for 
\eqref{44} and 
\eqref{45}.
 \end{proof}

By Theorem 1, the last lemma establishes the existence of an entropy 
function $S$ within the context of one
simple system $\Gamma$ and its scaled copies. Axiom A7 implies that 
$S$ is a {\it concave} function of $X=(U,V)\in \Gamma$, i.e.,
\begin{equation}\label{concaveS}
 (1-\lambda)S(U,V)+\lambda 
 S(U',  V')\leq 
 S((1-\lambda)U+\lambda U',  (1-\lambda)V+\lambda V').
 \end{equation}   
Moreover, by A11 and A12 and the properties of entropy described in 
Theorem 1 (i),
\begin{equation}\label{entropymax}
(U,V)\simt (U',V') \ \Longleftrightarrow \ S(U,V)+S(U',V')=\max_{W}[ 
 S(W,V)+S(U+U'-W,V')]. 
 \end{equation}     
For a given $\Gamma$ the entropy function is unique up to a 
mutiplicative and an additive constant which are indetermined as
long as we stay within the group
of scaled copies of $\Gamma$. The next task is to show that the
multiplicative constants can be adjusted to give a universal entropy
valid for copies of {\em different} systems, i.e. to establish the 
hypothesis of Theorem 2.  This is based on the
following.

\begin{lem}{Lemma 5 {\rm (Existence of calibrators)}}
If $\Gamma_1$ and
$\Gamma_2$ are simple systems, then there exist states
$X_0,X_1\in\Gamma_1$ and $Y_0,Y_1\in\Gamma_2$ such that
\begin{equation}
X_0\prec\prec X_1 \quad\quad {\rm and}\quad\quad Y_0\prec\prec Y_1
\end{equation}
and
\begin{equation}\label{calib}
(X_0,Y_1)\sima (X_1,Y_0)\,.
\end{equation}
\end{lem}

The significance of this lemma is that it allows us to fix  the 
{\em multiplicative} constants by the condition
\begin{equation}\label{calibration}
S_1(X_0) + S_2(Y_1) = S_1(X_1)+S_2(Y_0)\,.
\end{equation}
\begin{proof}[Proof of Lemma 5]
The proof of this lemma is not entirely simple and it involves all the
axioms A1--A15.  Consider the simple system $\Delta_{12}$ obtained by
thermally joining $\Gamma_1$ and $\Gamma_2$.  Let $Z$ be some
arbitrary point in $\Delta_{12}$ and consider the adiabat $\partial
A_{Z}$.  Any point in $\partial A_{Z} $ is by Axiom A12 adiabatically
equivalent to some pair $(X,Y)\in\Gamma_{1}\times\Gamma_{2}$.  There
are now two alternatives.
\begin{itemize}
    \item For some $Z$ there are two such pairs, $(X_{0}, Y_{1})$ and $(X_{1}, 
    Y_{0})$ such that $X_{0}\prec\prec X_{1}$. Since $(X_{0}, 
    Y_{1})\sima Z\sima (X_{1}, Y_{0})$, this implies 
    $Y_{0}\prec\prec Y_{1}$ and we are done.
    \item $(X,Y)\sima (\bar X,\bar Y)$ with $X\simt Y$ and $\bar X\simt 
    \bar Y$ always 
    implies $X\sima \bar X$ (and hence also $Y\sima \bar Y$).
    \end{itemize}
The task is thus to exclude the second alternative.

The second alternative is certainly excluded if the thermal splitting 
of some $Z=(U,V_{1},V_{2})$ in $ \Delta_{12}$ is not unique. Indeed, if 
$U=U_{1}+U_{2}=\bar U_{1}+\bar U_{2}$ with $U_{1}<\bar U_{1}$ and 
$Z\sima (X,Y)\sima (\bar X,\bar Y)$ with $X=(U_{1},V_{1})$, $\bar 
X=(\bar U_{1},V_{1})$,
$Y=(U_{2},V_{2})$ and $\bar Y=(\bar U_{2},V_{2})$, then
$X\prec\prec \bar X$ and $\bar Y\prec\prec Y$. Hence we may assume 
that for every $Z=(U,V_{1},V_{2})\in \Delta_{12}$ there are {\it 
unique} $U_{1}$ and $U_{2}$ with $U_{1}+U_{2}=U$ and
$Z\sima ((U_{1},V_{1}),(U_{2},V_{2}))$.

Consider now some fixed $\bar Z\in \Delta_{12}$ with corresponding
thermal splitting $(\bar X,\bar Y)$, $\bar X\in\Gamma_{1}$, $\bar Y\in
\Gamma_{2}$.  We shall now show that the second alternative above
leads to the conclusion that all points on the adiabat $\partial
A_{\bar Z}$ are in thermal equilibrium with each other.  By the zeroth
law (Axiom A13) and since the domain of work coordinates corresponding
to the adiabat $\partial A_{\bar Z}$ is connected (by Axiom A10), it
is sufficient to show this for all points with a fixed work coordinate
$V_{1}$ and all points with a fixed work coordinate $V_{2}$.

The second alternative means that if $Z\in \partial A_{\bar Z}$ has
the thermal splitting $(X,Y)$, then $X\sima \bar X$ and $Y\sima \bar
Y$.  For a given work coordinate $V_{1}$ there is a unique $\tilde
X=(\tilde U_{1},V_{1})\in\partial A_{\bar X}$.  (Its energy coordinate
is uniquely determined as a solution of the partial differential
equations for the adiabat, cf.\ Step 3 in the proof of Theorem 3.)
Hence the thermal splitting of each points $Z$ with fixed work
coordinate $V_{1}$ has the form $(\tilde X,Y)$, and $Y\simt \tilde X$
for all such $Y$.  By the zeroth law, all $Y$'s are in thermal
equilibrium with each other (since they are in thermal equilibrium
with a common $\tilde X$), and hence, by the zeroth law and Lemma 3,
all the points $\theta(\tilde X,Y)$ are in thermal equilibrium with
each other.  In the same way one shows that all points in $\partial
A_{\bar Z}$ with fixed $V_{2}$ are in thermal equilibrium with each
other.

To complete the proof we now show that if a simple system, in
particular $\Delta_{12}$, contains two points that are {\it not} in
thermal equilibrium with each other, then there is at least one {\it
adiabat} that contains such a pair.  The case that all points in
$\Delta_{12}$ are in thermal equilibrium with each other can be
excluded, since by A15 it would imply the same for $\Gamma_{1}$ and
$\Gamma_{2}$ and thus the thermal splitting would not be unique,
contrary to assumption.  (Note, however, that a world where all
systems are in thermal equilibrium with each other is not in conflict
with our axiom system.  The entropy would then be an affine fuction of
$U$ and $V$ for all systems.  In this case, the first alternative
above would always hold.)

In our proof of the existence of an adiabat with two points not in 
thermal equilibrium we shall make use of the already established 
entropy function $S$ for the simple system $\Gamma=\Delta_{12}$  
which characterizes the adiabats in $\Gamma$ and 
moreover has the properties \eqref{concaveS} and \eqref{entropymax}. 

The fact that $S$  characterizes the adiabats means that if $\sr
\subset \R$ denotes the range of $S$ on $\Gamma$ then the sets
\begin{equation}
E_\sigma = \{ X \in \Gamma : S(X) =
\sigma \}, \qquad \sigma \in \sr
\end{equation}
are precisely the adiabats of
$\Gamma$. Furthermore, the
concavity of $S$ --- and hence its continuity on the connected open set
$\Gamma$ --- implies that $\sr$ is connected, i.e., $\sr$ is an
interval.

Let us assume now that for any adiabat, all points on that adiabat are 
in thermal equilibrium with each other. We have to show that this 
implies  that all points in $\Gamma$ are in thermal equilibrium with 
each other.
By the zeroth law, A3,  and Lemma 2 (i), $\simt$ is an
equivalence relation that divides $\Gamma$ into disjoint equivalence
classes.  By our assumption,  each such equivalence class must be a
union
of adiabats, which means that the equivalence classes are represented
by a
family of disjoint subsets of $\sr$.  Thus
\begin{equation}
\sr = \bigcup \limits_{\alpha \in \I} \sr_\alpha
\end{equation}
where $\I$ is some index set, $\sr_\alpha$ is a subset of $\sr$,
$\sr_\alpha \cap \sr_\beta = 0$ for $\alpha \not= \beta$, and $E_\sigma
\simt E_\tau$ if and only if $\sigma$ and $\tau$ are in some common
$\sr_\alpha$.

We will now prove that each $\sr_\alpha$ is an open set.  It is then an
elementary topological fact (using the connectedness of $\Gamma$) 
that there can be
only one non-empty $\sr_\alpha$, i.e., all points in $\Gamma$ are in 
thermal equilibrium with each other and our proof will be 
complete.

The concavity of $S(U,V)$ with respect to $U$ for each fixed $V$
implies
the existence of an upper and lower $U$-derivative at each point, which we 
denote by 
$1/T_+ $ and $1/T_- $, i.e., 
\begin{equation}\label{74}
(1/T_\pm) (U,V) = \pm \lim \limits_{\varepsilon \searrow 0}
\varepsilon^{-1} [S(U \pm \varepsilon, V) - S(U,V)].
\end{equation}
Eq.\ \eqref{entropymax} implies that $X \simt Y$ if and only if the closed 
intervals
$[T_- (X), T_+ (X)]$ and $[T_- (Y), T_+ (Y)]$ are not disjoint.
Suppose
that some $\sr_\alpha$ is not open, i.e., there is $\sigma \in
\sr_\alpha$
and either a sequence $\sigma_1 > \sigma_2 > \sigma_3 \cdots$,
converging
to $\sigma$ or a sequence $\sigma_1 < \sigma_2 < \sigma_3 < \cdots$
converging to $\sigma$ with $\sigma_i \not\in \sr_\alpha$.  Suppose the
former (the other case is similar).  Then (since $T_\pm$ are
monotone
increasing in $U$ by the concavity of $S$) we can conclude that for
{\it
every} $Y \in E_{\sigma_i}$ and {\it every} $X \in E_\sigma$
\begin{equation}\label{4.10}
T_- (Y) > T_+ (X). 
\end{equation}
We also note, by the monotonicity of $T_\pm$ in $U$, that \eqref{4.10}
necessarily holds if
$Y \in E_\mu$ and $\mu \geq \sigma_i$; hence (1) holds for all $Y \in
E_\mu$ for {\it any} $\mu > \sigma$ (because $\sigma_i \searrow
\sigma$).  On the other hand, if $\tau \leq \sigma$
\begin{equation}
T_+ (Z) \leq T_-(X)
\end{equation}
for $Z \in E_\tau$ and $X \in E_\sigma$.  This contradicts
transversality, namely the hypothesis that there is $\tau < \sigma <
\mu$, $Z \in E_\tau, Y \in E_\mu$ such that $[T_- (Z), T_+ (Z)] \cap
[T_- (Y), T_+ (Y)]$ is not empty. 
\end{proof}

With  the aid of Lemma 5 we now arrive at our chief goal, which is CH for 
compound systems. 

\begin{thm}{Theorem 4 {\bf (Entropy principle in products of simple
systems)}}
The comparison hypothesis CH is valid in arbitrary  compounds of
simple systems. Hence, by Theorem 2, the relation $\prec$ among states
in such state-spaces is characterized by an entropy function $S$. The
entropy function is \underbar {unique}, up to an overall multiplicative
constant and one additive constant for each simple system under
consideration.
\end{thm}
\begin{proof}
Let $\Gamma_{1}$ and    $\Gamma_{2}$ be simple systems and let $X_{0}, 
X_{1}\in\Gamma_{1}$ and  $Y_{0}, 
Y_{1}\in\Gamma_{2}$ be points with the properties described in Lemma 5. 
 By Theorem 1 we  know that for every 
$X\in\Gamma_{1}$ and $Y\in\Gamma_{2}$
\begin{equation}X\sima ((1-\lambda_{1})X_{0},\lambda_{1} 
X_{1})\quad\hbox{and}\quad Y\sima ((1-\lambda_{2})Y_{0},\lambda_{2} 
Y_{1})\end{equation}
for some $\lambda_{1}$ and $\lambda_{2}$.
Define $Z_{0}=(X_{0},Y_{0})$ and $Z_{1}=(X_{1},Y_{1})$.
It is then simple arithemtics, making use of \eqref{calib} besides 
Axioms A3--A5, to show 
that
\begin{equation}(X,Y)\sima((1-\lambda)Z_{0},\lambda 
Z_{1})\end{equation} with
$\lambda=\mfr1/2(\lambda_{1}+\lambda_{2})$. By the equivalence of
\eqref{refpointS} and \eqref{lambdalambda} we know that this is
sufficient for comparability within the state space $\Gamma_{1}\times
\Gamma_{2}$. 

Consider now a third simple system $\Gamma_{3}$ and apply Lemma 3
to $\Delta_{12}\times \Gamma_{3}$, where 
$\Delta_{12}$ is the thermal join of $\Gamma_{1}$ and 
$\Gamma_{2}$. By Axiom A12 the reference points in  $\Delta_{12}$ are 
adiabatically equivalent to points in $\Gamma_{1}\times \Gamma_{2}$, 
so we can repeat the reasoning above and conclude that all points 
in $(\Gamma_{1}\times \Gamma_{2})\times\Gamma_{3}$ are comparable. 
By induction, this extends to an arbitrary products of simple systems. 
This includes multiple scaled products, because by Lemma 
3 and Theorem 1, every state in a multiple scaled product of copies of a 
simple 
system $\Gamma$ is adiabatically equivalent to a state in a 
single scaled copy 
of $\Gamma$.
\end{proof} 

\begin{rem}{remark}It should be emphasized that Theorem 4 contains more than
  the Entropy Principle for compounds of simple
  systems. The core of the theorem is an assertion about the {\em
    comparability} of all states in any state space composed of simple 
systems. (Note that the entropy principle would
  trivially be true if no state was comparable to any other state.)
  Combining Lemma 5 and Theorem 4 we can even assert that certain
  compound states in {\em different} state spaces are comparable: What
  counts is that the total \lq mass\rq\ of each simple system that
  enters the compound is the same for both states.  For instance, if
  $\Gamma_1$ and $\Gamma_2$ are two simple systems, and
  $X_1,\dots,X_N,X'_1,\dots,X'_{N'}\in \Gamma_1$, 
$Y_1,\dots,Y_M,Y'_1,\dots,Y'_{M'}\in \Gamma_2$, then
$Z=(\lambda_1X_1,\dots,\lambda_NX_N, \mu_1Y_1,\dots,\mu_MY_M)$  is 
comparable to  $W=(\lambda'_1X'_1,\dots,\lambda'_{N'}X'_{N'}, 
\mu'_1Y'_1,\dots,\mu'_{M'}Y'_{M'})$ provided $\sum\lambda_i=\sum\lambda'_j$,
 $\sum\mu_k=\sum \mu'_\ell$, and in this case $Z\prec W$ if and only if 
$S(Z)\leq S(W)$.  
\end{rem}
At last, we are now ready to  define  {\em temperature}.  Concavity of
$S$  (implied by A7),  Lipschitz
continuity of the pressure and the transversality condition, together
with some real analysis, play key roles in the following, which answers
questions Q3 and Q4 posed at the beginning.

\begin{thm}{Theorem 5 {\bf(Entropy defines temperature)}}
The entropy, $S$, is a concave and continuously differentiable
function on the state space of a simple system.  If the function $T$
is defined by
\begin{equation}
\tfrac1T:=\left(\tfrac{\partial S}{ \partial U}\right)_V
\end{equation}
then $T > 0 $ and $T$ characterizes the relation $\simt$ in the sense
that $X\simt Y$ if and only if $T(X)=T(Y)$.  Moreover, if two systems
are brought into thermal contact with fixed work coordinates then,
since the total entropy cannot decrease, the energy flows from the
system with the higher $T$ to the system with the lower $T$.
\end{thm}

\begin{rem}{Remark}
The temperature need not be a strictly monotone function of $U$; indeed,
it is not so in a `multiphase region' (see Fig.\ 5). It follows  that $T$ is not
always  capable of specifying  a state, and this fact  can cause some
pain in traditional discussions of the second law -- if
it is recognized, which usually it is not. 
\end{rem}

\begin{figure}[htq]\begin{center}
      \includegraphics[width=10cm]{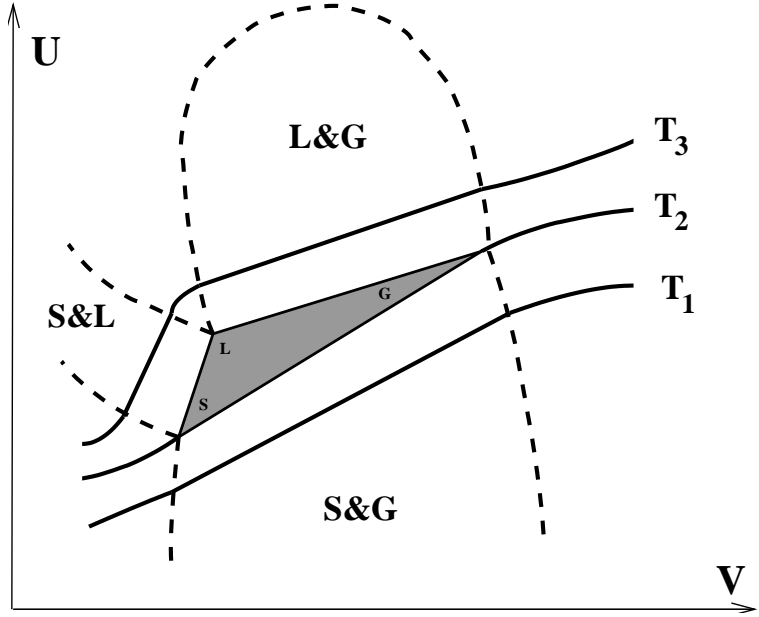}\end{center}
\caption{Isotherms in the (U,V) plane near the triple point 
  (L=liquid, G=gas, S=solid) of a simple system.  (Not to scale). In
  the triple point region the temperature is constant, which shows
  that an isotherm need not have codimension one.}
\label{figure5_cdm}
\end{figure}


\begin{proof}[Proof of Theorem 5.]
  The complete proof is rather long and we shall not bring all details 
 here. They can be found in \cite{[7]} (Lemma 5.1 
  and Theorems 5.1--5.4 .) 
 As in the proof of Lemma 5, concavity of $S(U,V)$ implies the 
 existence of the upper and lower partial derivatives of $S$ with 
 respect to $U$ and hence of the upper and lower temperatures 
 $T_{\pm}$ defined by \eqref{74}. Moreover, as also noted in the 
 proof of Lemma 5, \begin{equation}\label{tequilibr}
 X\simt Y\quad\Longleftrightarrow\quad [T_{-}(X), 
 T_{+}(X)]\cap[T_{-}(Y), T_{+}(Y)]\neq \emptyset.
 \end{equation}
 The main goal is to show that $T_{-}=T_{+}=T$, and that $T$ is 
 a continuous function of $X$.
 
 {\it Step 1:  $T_{+}$ and $T_{-}$ are locally Lipschitz 
 continuous on adiabats.} The essential input here is the local 
 Lipschitz continuity of the 
pressure,  i.e., for each $X\in\Gamma$ and each $r>0$ there is a constant 
 $C=C(X,r)$ such that
 \begin{equation}\label{lcp} |P(X)-P(Y)|\leq C\,|X-Y|
  \end{equation}   
if $|X-Y|<r$. The assertion is that
 \begin{equation}\label{lct} |T_{+}(X)-T_{+}(Y)|\leq c\,|X-Y|
  \end{equation}   
 for some $c=c(X,r)$, if  $|X-Y|<r$ and $Y\in\partial A_{X}$, 
 together with the analogous 
 equation for $T_{-}$).

As in the proof of Theorem 3, Step 3, the adiabatic surface through
$X=(U_{0},V_{0})$ is given by $(u_{X}(V),V)$ where $u_{X}$ is the
solution to the system of differential equations
\begin{equation}\label{adiabateq}
    \frac{\partial u}{\partial V_{i}}=-P_{i},\quad i=1,\dots,n
    \end{equation}
    with the inital condition $u(V_{0})=U_{0}$.  Let us denote this
    solution by $u_{0}$, and consider for $\varepsilon>0$ also the
    solution $u_{\varepsilon}$ with the initial condition
    $u(V_{0})=U_{0}+\varepsilon$.  This latter solution determines
     the adiabatic surface through
    $X_{\varepsilon}=(U_{0}+\varepsilon,V_{0})$ (for $V$ sufficiently close to $V_{0}$, so that
    $(u(V),V)\in\Gamma$).
    
    Let $S_{0}$ denote the entropy on $(u_{0}(V),V)$ and 
    $S_{\varepsilon}$ the entropy on $(u_{\varepsilon}(V),V)$. Then, 
    by definition,
    \begin{equation}
        T_{+}(U_{0},V_{0})=\lim_{\varepsilon\downarrow 0}
        \frac{\varepsilon}{S_{\varepsilon}-S_{0}}=
        \lim_{\varepsilon\downarrow 0
        }\frac{u_{\varepsilon}(V_{0})-u_{0}(V_{0})}
        {S_{\varepsilon}-S_{0}}
    \end{equation}
 and 
 \begin{equation}
        T_{+}(u_{0}(V),V)=
        \lim_{\varepsilon\downarrow 0}\frac{u_{\varepsilon}(V)-u_{0}(V)}
        {S_{\varepsilon}-S_{0}}=T_{+}(U_{0},V_{0})\left[1+
        \lim_{\varepsilon\downarrow 0}
        \frac{u_{\varepsilon}(V)-u_{0}(V)-
        \varepsilon}
        {\varepsilon}\right].
        \end{equation}
To prove \eqref{lct} it suffices to show that for all nonnegative 
$\varepsilon$ 
close to 0 and $V$ close to $V_{0}$
\begin{equation}
\frac{u_{\varepsilon}(V)-u_{0}(V)-
        \varepsilon}
        {\varepsilon}\leq D\,|V-V_{0}|.
\end{equation}
for some $D$. This estimate (with $D=2C$) follows from 
\eqref{lcp}, using \eqref{adiabateq} to write 
$u_{\varepsilon}(V)-u_{\varepsilon}(V_{0})$ and $u_{0}(V)-u_{0}(_{0})$ 
as line integrals of the 
pressure. See \cite{[7]}, p. 69.
 
 {\it Step 2: $T_{+}(X)>T_{-}(X)\,\Rightarrow\,T_{+}$ and $T_{-}$ are 
 constant on $\partial A_{X}$.}
 This step relies on concavity of entropy, continuity of $T_{\pm}$ on 
 adiabats (Step 1), and last but not least, on the zeroth law. Without 
 the zeroth law it is easy to give counterexamples to the assertion.
 
 If $T_{+}(X)>T_{-}(X)$, but $T_{+}$ is not constant, then by continuity of 
 $T_{\pm}$ on adiabats 
 there exist $Y,Z$ with $Y\sima 
 Z\sima X$, $T_{+}(Y)<T_{+}(Z)$ but $[T_{-}(Y), 
 T_{+}(Y)]\cap[T_{-}(Z), T_{+}(Z)]\neq \emptyset$, i.e., $Y\simt Z$.
 Now it is a general fact about a concave function, 
 in particular $U\mapsto S(U,V)$, that the set of points where it 
 is differentiable, i.e., where $T_{+}(U,V)=T_{-}(U,V)\equiv 
 T(U,V)$, is 
 dense. Moreover, if $U_{1}>U_{2}>\dots$ is a sequence of such points 
 converging to $U$, then $T(U_{i},V)$ converges to $T_{+}(U,V )$.
 Using continuity of $T_{+}$ on the adiabat, we conclude that 
 there exists a $W$ such that $T_{+}(W)=T_{-}(W)=T(W)$, but
 $T_{+}(Y)<T(W)<T_{+}(Z)$. This contradicts the zeroth law, because 
 such a $W$ would be in thermal equilibrium with $Z$ (because 
 $T(W)\in[T_{-}(Z), T_{+}(Z)]$) but not with $Y$ (because $T(Y)\notin
 [T_{-}(Y), T_{+}(Y)]$). In the same way one leads the assumption 
 that $T_{-}$ is not constant on the adiabat $\partial A_{X}$ to a 
 contradiction.
 
 {\it  Step 3: $T_{+}=T_{-}.$} Assume $T_{+}(X)>T_{-}(X)$ for some 
 $X$. Then, by Step 2, $T_{+}$ and $T_{-}$ are constant on the whole 
 adiabat $\partial A_{X}$. Now by concavity and monotonicity of $S$ in 
 $U$ (cf.\ the remark following the proof of Theorem 3) we have 
 $T_{+}(Y)\leq T_{-}(Z)$  if $S(U_{Y},V_{Y})< S(U_{Z},V_{Z})$, 
 $V_{Y}=V_{Z}$. Hence, 
 if $Z$ is such that \begin{equation}\label{z}
 Y\prec\prec Z\ \hbox{and $V_{Y}=V_{Z}$ for some 
 $Y\sima X$},
 \end{equation}then
 $T_{+}(X)\leq T_{-}(Z)\leq T_{+}(Z)$.  Likewise, if $Z'$ is such that 
 \begin{equation}\label{z'} Z'\prec\prec Y'\ 
     \hbox{ and $V_{Y'}=V_{Z'}$ for some 
 $Y'\sima X$},
 \end{equation}then 
 \begin{equation}\label{gap}T_{-}(Z')\leq T_{+}(Z')\leq T_{-}(X)<T_{+}(X)\leq T_{-}(Z)
     \leq T_{+}(Z).
  \end{equation}
  This means that no $Z$ satisfying \eqref{z} can be in thermal 
  equilibrium with a $Z'$ satisfying \eqref{z'}. In the case that 
  every point in the state space has its work coordinates in common 
  with some point on the adiabat $\partial A_{X}$, this violates the transversality 
  axiom, A14. 
  
  Using axiom A5 we can also treat the case where the projection of the 
  adiabat $\partial X_{X}$ onto the work coordinates, 
  does 
  not cover the whole range of the work coordinates for $\Gamma$, 
  i.e., when
  \begin{equation}
   \rho_{X}:=\{V:\ (U,V)\in  \partial A_{X}\  \hbox{for some}\ U\}\neq
  \{V:\ (U,V)\in  \Gamma\ \hbox{for some}\ U\}=: \rho(\Gamma).
  \end{equation}
  One 
  considers a line of points $(U,\bar V)\in\Gamma$ with $\bar V$ fixed
  on the boundary of $\rho_{X}$ in $\rho(\Gamma)$.
  One then shows that a gap between the upper and lower temperature 
  of $X$, i.e., $T_{-}(X)<T_{+}(X)$, implies that points 
  with these work coordinates $\bar V$ can only be in thermal 
  equilibrium with points on one side of the adiabat $\partial 
  A_{X}$, in contradiction to A15 and the zeroth law. 
  See \cite{[7]}, p. 72 for the details. 
  
 {\it Step 4: $T$ is continuous.} By Step 3 $T$ is uniquely defined 
 and by Step 1 it is locally Lipschitz continuous on each adibat, i.e.,
 \begin{equation}\label{cont}
    |T(X)-T(X')|\leq c\,|X-X'] 
 \end{equation}    
 if $X\sima X'$ and $X$ and $Y$ both lie in some ball of sufficiently small 
 radius.  Moreover, 
 concavity of $S$ and the fact that $T_{+}=T_{-}=T$ imply that 
 $T$ is continuous along each line $l_{V}=\{(U,V):\ 
 (U,V)\in\Gamma\}$.
 
 Let now $X_\infty, X_1, X_2, \dots$ be points in $\Gamma$
such that $X_j \rightarrow X_\infty$ as $j \rightarrow \infty$.  We
write $X_j = (U_j, V_j)$, we let $A_j$ denote the adiabat $\partial
A_{X_j}$, we let $T_j = T(X_j)$ and we set $l_j = \{ (U, V_j): (U, V_j)
\in \Gamma \}$. 

By axiom A9, the slope of the tangent of $A_{X}$, i.e., 
the pressure $P(X)$, is locally Lipschitz continuous.
Therefore for $X_{j}$ sufficiently close to $X_\infty$ we can
assume that each adiabat $A_j$ intersects $l_\infty$ in some point,
which we denote by $Y_j$.  Since $\vert X_j - X_\infty \vert
\rightarrow 0$ as $j \rightarrow \infty$, we have that $\vert Y_j
\rightarrow X_\infty \vert$ as well.  In particular, we can assume 
that all the $X_{i}$ and $Y_j$ lie in some small ball around 
$X_{\infty}$ so that \eqref{cont} applies. Now
\begin{equation}
\vert T(X_j) -
T(X_\infty) \vert \leq \vert T(X_j) - T(Y_j) \vert + \vert T(Y_j) -
T(X_\infty) \vert,\end{equation}
and as $j \rightarrow \infty$, $T(Y_j) - T(X_\infty)
\rightarrow 0$ because $Y_j$ and $X_\infty$ are in $l_\infty$.  Also,
$T(X_j) - T(Y_j) \rightarrow 0$ because $\vert T(X_j) - T(Y_j) \vert <
c \vert X_j - Y_j \vert \leq c \vert X_j - X_\infty
\vert + c \vert Y_j - X_\infty \vert$.  

 {\it Step 5: $S$ is continuously differentiable.} The adiabat through a point 
 $X \in \Gamma$ is characterized
by the once continuously differentiable function, $u_X (V)$, on $\R^n$. 
Thus, $S(u_X (V), V)$ is constant, so (in the sense of distributions)
\begin{equation}
0 = \left(\frac {\partial S }{\partial U} \right) \left( \frac{\partial
u_X}
{ \partial V_j} \right) + \frac{\partial S }{ \partial V_j}.
\end{equation}
Since $1/T = \partial S /\partial U$ is continuous, and $\partial u_X
/\partial V_j = -P_j$ is Lipschitz continuous, we see that $\partial
S/\partial V_j$ is a continuous function
and we have the well known formula
\begin{equation}
\frac{\partial S }{{\partial V_j }}= \frac{P_j }{ T}.
\end{equation}
 
 {\it Step 6: Energy flows from `hot' to `cold'.} Let 
 $X=(U_{X},V_{X})$ and $Y=(U_{Y},V_{Y})$ be two states of simple 
 systems and assume that $T(X)>T(Y)$. By Axioms A11 and A12
 \begin{equation}\label{flow}
  (X,Y)\prec \theta(X,Y)\sima (X',Y')   
 \end{equation} 
 with $X'=(U_{X'},V_{X})$, $Y'=(U_{Y'},V_{Y})$ and 
 \begin{equation}\label{conserv}
  U_{X'}+U_{Y'}=U_{X}+U_{Y}.
 \end{equation} 
  Moreover, $X'\simt Y'$ and hence, by \eqref{tequilibr} and Step 3
  \begin{equation}\label{tempeq}
  T(X')=T(Y')\equiv T^{*}.
   \end{equation} 
   We claim that
\begin{equation}\label{middle}
T(X) \geq T^{*} \geq T(Y) .  
 \end{equation} 
(At least one of 
these inequalities is strict because of the uniqueness of temperature for
each state.)  Suppose that inequality \eqref{middle} failed, e.g., 
$T* > T(X) >T(Y)$. Then we would have that 
$U_{X'} > U_{X}$ and $ U_{Y'} > U_{Y}$ and at least one of these would be strict
(by the strict monotonicity of $U$ with respect to  $T$, 
which follows from the concavity and differentiability of $S$). 
This pair of inequalities is impossible in view of \eqref{conserv}.

Since $T^{*}$ satisfies \eqref{middle}, the theorem now follows 
from the monotonicity of $U$ with respect to $T$. 
 \end{proof}
 
 {}From the entropy principle and the relation 
\begin{equation}T=(\partial S/\partial U)^{-1}\end{equation}
between temperature and entropy 
we can now derive the usual formula for the {\bf Carnot efficiency} 
\begin{equation}\label{(5.5)}\eta_{\rm C}:=1-(T_0/T_1)
\end{equation}
as 
an upper bound for the efficiency of a `heat engine' that undergoes a 
cyclic process. Let us define a {\bf thermal
reservoir} to be a simple system whose work coordinates remains unchanged
during some process. Consider a combined system consisting of a 
thermal reservoir and
some machine, and an adiabatic process for this combined system.  
The entropy principle says that the total entropy change in this process 
is 
\begin{equation}\label{(5.6)}\Delta S_{\rm machine}+\Delta
S_{\rm reservoir}\geq 0.\end{equation}  Let $-Q$ be the energy change of the 
reservoir,
i.e., if $Q\geq 0$, then the reservoir delivers energy, otherwise it
absorbs energy. If $T$ denotes the temperature of the reservoir {\it at 
the
end of the process}, then, by the convexity of $S_{\rm reservoir}$ in $U$, 
we have
\begin{equation}\label{(5.7)}
    \Delta S_{\rm reservoir}\leq -\frac{Q}{ T}.\end{equation} Hence 
\begin{equation}\label{(5.8)}\Delta S_{\rm machine}-\frac{Q}{
T}\geq 0.\end{equation} 
Let us now couple the machine first to a \lq\lq high temperature
reservoir\rq\rq\ which delivers energy $Q_{1}$ and reaches a final temperature 
$T_1$,
and later to a \lq\lq low 
temperature
reservoir\rq\rq\ which absorbs energy $-Q_{0}$ and reaches a final temperature 
$T_0$. The whole process is 
assumed 
to be cyclic for the machine so the entropy changes for the machine in 
both steps cancel. (It
returns to its initial state.) Combining \eqref{(5.6)}, \eqref{(5.7)} and 
\eqref{(5.8)} we 
obtain 
\begin{equation}\label{(5.9)}Q_1/T_1+Q_0/T_0\leq 0
\end{equation} 
which gives the usual inequality for the efficiency $\eta := 
(Q_{1}+Q_{0})/Q_{1}$:
\begin{equation}\label{(5.10)}
    \eta\leq 1-(T_0/T_1)=\eta_{\rm C}.\end{equation} 
In text book presentations it is usually assumed that the reservoirs are
infinitely large, so that their temperature remains unchanged, but formula
\eqref{(5.10)} remains valid for finite reservoirs, provided $T_1$ and $T_0$ are
properly interpreted, as above.  
    
\subsection*{Mixing and chemical reactions.}
The core results of our 
analysis have now been presented and readers satisfied with the 
entropy principle in the form of Theorem 4 may wish to stop
at this point.  Nevertheless, a
nagging doubt will occur to some, because there are important adiabatic
processes in which systems are not conserved, and these processes are
not yet covered in the theory. A critical 
study of the usual texbook treatments should
convince the reader that this subject is not easy, but
in view of the manifold applications of thermodynamics to chemistry and 
biology it is important to tell the whole story
and not  ignore such processes.

One can formulate the problem as the determination of 
the additive constants
$B{(\Gamma)}$ of Theorem 2. As long as we consider only adiabatic
processes that preserve the amount  of each simple system (i.e., such
that Eqs. (6) and (8) hold), these constants are indeterminate. This is
no longer the case, however, if we consider mixing processes and
chemical reactions (which are not really different, as far as
thermodynamics is concerned.) It then becomes a nontrivial question
whether the additive constants can be chosen in such a way that the
entropy principle holds. Oddly, this determination turns out to be far
more complex, mathematically and physically  than the determination of
the multiplicative constants (Theorem 2). In
traditional treatments one resorts to {\em gedanken} experiments
involving idealized devices such as `van t'Hofft boxes' which are made
of idealized materials such as `semipermeable membranes' that do not
exist in the real world except in an approximate sense in a few
cases. For the derivation of the entropy principle by this method, 
however, one needs
virtually perfect `semipermeable membranes' for all substances, and it
is fair to question whether such a precise physical law should be
founded on non-existent objects. Fermi, in his famous textbook
\cite{F}, draws attention to this problem, but, like those before
him and those after him, chooses to ignore it and presses on. We
propose a better way. 

What we already know is that every system has a 
well-defined entropy function, e.g., for each  $\Gamma$ 
there is $S_\Gamma$, and we know from Theorems 2 and 4 that the multiplicative 
constants $a_{\Gamma}$ can been
determined
in such a way that the sum of the entropies increases in any  adiabatic
process in any compound space
$\Gamma_1 \times  \Gamma_2 \times ...$. Thus, if
$X_i \in  \Gamma_i$ and $Y_i \in  \Gamma_i$ then
\begin{equation}\label{(20)}
(X_1,X_2,...) \prec (Y_1,Y_2,...) \quad {\rm if \ and \ only \ if} \quad
\hbox{$\sum_{i}$}S_i(X_i)\leq \hbox{$\sum_{j}$}S_j(Y_j)\,. \end{equation}
where we have denoted $S_{\Gamma_i}$ by $S_i$ for short.  The additive
entropy constants do not matter here since each function $S_i$ appears
on both sides of this inequality. It is important to note that this
applies even to processes  that, in intermediate steps,  take one system
into another, provided the total compound system is the same at the
beginning and at the end of the process.  

The task is to find 
constants $B(\Gamma)$, one for each 
state space $\Gamma$, in such a way that the 
entropy defined by
\begin{equation}\label{(21)}
S(X) := S_\Gamma(X) + B(\Gamma)  \quad\quad {\rm for} \quad\quad X \in 
\Gamma    
\end{equation}\label{(22)}
satisfies
\begin{equation}
S(X) \leq S(Y)   
\end{equation}
whenever 
\begin{equation}
X\prec Y  \quad\quad {\rm with } \quad\quad X \in 
\Gamma\,, \ Y \in \Gamma'\,. 
\end{equation}
Additionally, we require that the  newly defined entropy satisfies
scaling and additivity under composition.
Since the initial entropies $S_\Gamma(X)$ already satisfy
them, these requirements become conditions on the
additive constants $B(\Gamma)$:
\begin{equation}\label{(23)}
B(\Gamma_1^{(\lambda_1)}\times\Gamma_2^{(\lambda_2)})=
\lambda_1B(\Gamma_1)+\lambda_2 B(\Gamma_2)
\end{equation}
for all state spaces $\Gamma_1,\Gamma_2$ under
considerations and $\lambda_1,\lambda_2>0$.
Some reflection shows us that consistency in the definition of the
entropy constants $B(\Gamma)$ requires us to consider all possible
chains of adiabatic processes leading from one space to another via
intermediate steps.  Moreover, the additivity requirement leads us to
allow the use of a `catalyst' in these processes, i.e., an auxiliary
system, that is recovered at the end, although a state change {\em
within} this system might take place.  With this in mind we define
quantities $F(\Gamma,\Gamma')$ that incorporate the entropy differences
in all such chains leading from $\Gamma$ to $\Gamma'$.  These are built
up from simpler quantities $D(\Gamma,\Gamma')$, which measure the
entropy differences in one-step processes, and $E(\Gamma,\Gamma')$,
where the `catalyst' is absent.
The precise definitions are as follows.
First, 
\begin{equation}\label{(24)}
D(\Gamma,\Gamma'):=\inf\big\{S_{\Gamma'}(Y)-S_{\Gamma}(X):
X\in\Gamma\,,\ Y\in\Gamma'\,,\ X \prec Y\big \}\,.
\end{equation}
If there is no adiabatic process leading from $\Gamma$ to $\Gamma'$ we
put $D(\Gamma,\Gamma')=\infty$. Next, 
for any
given $\Gamma$ and $\Gamma'$ we consider all 
finite chains of state spaces, 
$\Gamma=\Gamma_1,\Gamma_2,\dots,\Gamma_N=\Gamma'$
such that $D(\Gamma_i,\Gamma_{i+1})<\infty$ for all i, and we define
\begin{equation}\label{(25)}
E(\Gamma,\Gamma'):=\inf\big\{D(\Gamma_1,\Gamma_{2})+
\cdots +D(\Gamma_{N-1},\Gamma_{N})
\big\}\,, 
\end{equation}
where the infimum is taken over all such chains linking $\Gamma$ with 
$\Gamma'$. Finally
we define
\begin{equation}\label{(26)}
F(\Gamma,\Gamma'):=\inf\big\{E(\Gamma\times\Gamma_0, \Gamma'\times
\Gamma_0)\big\} \,, 
\end{equation}
where the infimum is taken over all state spaces $\Gamma_0$. (These 
are the `catalysts'.)

The definition of the constants $F(\Gamma,\Gamma')$ involves a
threefold infimum and may look somewhat complicated at this point. 
The $F$'s, however, possess  subadditivity and invariance 
properties that need not
hold for the $D$'s and $E$'s, but are essential for an application of 
the Hahn-Banach theorem in the proof of
Theorem 7 below.  The importance of the $F$'s for the problem of the 
additive constants is made clear by the following theorem.

\begin{thm}{Theorem 6 {\bf(Constant entropy differences)}}
If $\Gamma$ and  $\Gamma'$ are two state spaces 
then for any two states $X\in \Gamma$ and  $ Y\in \Gamma'$ 
\begin{equation}\label{(27)}
X\prec Y \quad \hbox{\rm if and only if} \quad S_\Gamma(X) +F(\Gamma,
\Gamma') ~\leq ~S_{\Gamma'}(Y)\,. 
\end{equation}
\end{thm}
\begin{rem} {Remark} Since $F(\Gamma,\Gamma')\leq D(\Gamma,\Gamma')$
the theorem is trivially true when  $F(\Gamma,\Gamma')=+ \infty$, 
in the sense that there is then no adiabatic 
process from $\Gamma$ to $\Gamma'$. The reason  for
the title `constant entropy differences' is that the minimum jump between 
the
entropies $S_\Gamma(X)$ and $S_{\Gamma'}(Y)$ for $X\prec Y$ to be possible 
is independent of $X$. An essential ingredient for the proof of this theorem 
is Eq.\ \eqref{(20)}. 
\end{rem}

\begin{proof}[Proof of Theorem 6]
The `only if' part is obvious because $F(\Gamma,\Gamma')\leq
D(\Gamma,\Gamma')$.  For the proof of the `if' part we shall for
simplicity assume that the infima in \eqref{(24)}, \eqref{(25)} and
\eqref{(26)} are minima, i.e., that they are obtained for some chain
of spaces and some states in these spaces.  The general case can be
treated very similarly by approximation, using
the stability axiom, A6.

We thus assume that
\begin{equation}
F(\Gamma, \Gamma')= D(\Gamma \times \Gamma_0, \Gamma_1)+ 
D(\Gamma_1, 
\Gamma_2)
+\cdots +D(\Gamma_N,\Gamma' \times \Gamma_0)
\end{equation}
for some state spaces $\Gamma_0$, $\Gamma_1$, 
$\Gamma_2$,...,
$\Gamma_N$ and that
\begin{eqnarray}
 D(\Gamma \times \Gamma_0, \Gamma_1)&=&S(Y_{1})-S_{\Gamma \times 
 \Gamma_0}(\tilde X,X_{0}) = S(Y_{1})-S_{\Gamma}(\tilde X) -
 S_{\Gamma_{0}}(X_{0})\\
 D(\Gamma_{i}, \Gamma_{i+1})&=&
 S_{\Gamma_{i+1}}(Y_{i+1})-S_{\Gamma_{i}}(X_{i})\\
 D(\Gamma_N,\Gamma' \times 
 \Gamma_0)&=&S_{\Gamma'\times\Gamma_{0}}(\tilde 
 Y,Y_{0})-S_{\Gamma_{N}}(X_{N})=S_{\Gamma'}(\tilde 
 Y)+S_{\Gamma_{0}}(Y_{0})-S_{\Gamma_{N}}(X_{N})
    \end{eqnarray}
for states $X_i \in \Gamma_i$ and $Y_i \in  \Gamma_i$, for
$i=0,...,N$, $\tilde X\in \Gamma $ and $\tilde Y\in \Gamma'$ with
\begin{equation}\label{xy}
(\tilde X, X_0) \prec Y_1,\quad 
X_i \prec Y_{i+1}\  \hbox{ for $i=1,...,N-1$} ,\quad
X_N \prec (\tilde Y, Y_0) .
\end{equation}
Hence
\begin{equation}\label{F}
F(\Gamma, \Gamma')= S_{\Gamma'}(\tilde Y) +
\sum_{j=0}^N S_{\Gamma_{j}}(Y_j) - 
S_{\Gamma}(\tilde X) -\sum_{j=0}^N S_{\Gamma_{j}}(X_j).
\end{equation}
{F}rom the assumed inequality $S_\Gamma(X) +F(\Gamma,\Gamma')\leq 
S_{\Gamma'}(Y)$ and \eqref{F} we conclude that
\begin{equation}
S_{\Gamma}(X)+S_{\Gamma'}(\tilde Y) +\sum_{j=0}^N S_{\Gamma_{j}}(Y_j) \leq 
S_{\Gamma}(\tilde X)+S_{\Gamma'}(Y) +\sum_{j=0}^N S_{\Gamma_{j}}(X_j). 
\end{equation}
However, both sides of this inequality can be thought of as the entropy of 
a state in the compound space $\hat \Gamma := \Gamma \times \Gamma' 
\times
\Gamma_0 \times \Gamma_1 \times \cdots \times \Gamma_N$. 
The entropy
principle \eqref{(20)} for $\hat \Gamma$ then tell us that
\begin{equation}\label{xxz}
(X, \tilde Y, Y_0,\dots ,Y_N) \prec (\tilde X, Y, X_0,\dots ,X_N).
\end{equation}
On the other hand, using \eqref{xy} and the axiom A3, we have 
that
\begin{equation}\label{xyz}
(\tilde X, X_0, X_1, ..., X_N) \prec (\tilde Y, Y_0, Y_1, ..., Y_N).
\end{equation}
(The left side is here in $\Gamma\times\Gamma_0 \times 
\Gamma_1 \times \cdots \times \Gamma_N$ and the right side in 
$\Gamma'\times\Gamma_0 \times \Gamma_1 \times \cdots \times \Gamma_N$.)
By A3 again, we  have from \eqref{xyz} that 
\begin{equation}
    (\tilde X, Y, X_0,\cdots ,X_N)\prec (Y,\tilde Y, Y_0, Y_1, ..., Y_N).
 \end{equation}
(Left side in $\Gamma\times\Gamma'\times\Gamma_0 \times 
\Gamma_1 \times \cdots \times \Gamma_N$, right side in 
$\Gamma'\times\Gamma'\times\Gamma_0 \times \Gamma_1 \times \cdots \times 
\Gamma_N$.)
 {}From \eqref{xxz} and transitivity of the relation $\prec$ we then
have 
\begin{equation}
(X, \tilde Y, Y_0, Y_1, ..., Y_N) \prec (Y,\tilde Y, Y_0, Y_1, ..., Y_N),
\end{equation}
and the desired conclusion, $X\prec Y$, follows from the cancellation law 
\eqref{cancellation}.
\end{proof}

According to Theorem 6 the  determination of the entropy 
constants $B(\Gamma)$ amounts to satisfying the inequalities
\begin{equation}\label{(28)}
-F(\Gamma',\Gamma)~\leq ~B(\Gamma)-B(\Gamma')~\leq~ F(\Gamma,\Gamma')
\end{equation}
together with the linearity condition \eqref{(23)}. It is clear that 
\eqref{(28)} can
only be satisfied with finite constants $B(\Gamma)$ and $B(\Gamma')$, if
$F(\Gamma,\Gamma')>-\infty$. To exclude the pathological case 
$F(\Gamma,\Gamma')=-\infty$ we
introduce our last  axiom A16, whose statement requires  the following
definition. 

\begin{rem}{Definition}
A state-space, $\Gamma$ is said to be {\em connected} to
another state-space $\Gamma'$  if there are states $X\in \Gamma$ and
$Y\in\Gamma'$, and state spaces $\Gamma_1,\dots,\Gamma_N$ with states
$X_i, Y_i\in\Gamma_i$, $i=1,\dots,N$, and a state space $\Gamma_0$ with
states $X_0,Y_0\in\Gamma_0$, such that 
\begin{equation}
(X,X_0)\prec Y_1\,,\quad X_i\prec Y_{i+1}\,,\quad i=1,\dots, N-1\,,
\quad X_N\prec (Y,Y_0)\,.
\end{equation} 

\begin{enumerate}
\item[\bf A16.] {\bf Absence of sinks.} If $\Gamma$
is connected to $\Gamma'$ then $\Gamma'$ is connected 
\hbox{to $\Gamma$}.
\end{enumerate}
\end{rem}

This axiom excludes $F(\Gamma,\Gamma')=-\infty$ because, on
general grounds,  one   always has 
\begin{equation}\label{(29)}
-F(\Gamma',\Gamma)\leq F(\Gamma,\Gamma') \,.      
\end{equation}
(See below.) Hence $F(\Gamma,\Gamma')=-\infty$ (which means, in
particular, that $\Gamma$ is connected \hbox{to
$\Gamma'$)} would imply $F(\Gamma',\Gamma)=\infty$, i.e.,
that there is no way back from $\Gamma'$ \hbox{to
$\Gamma$}.
This is excluded by Axiom 16.

The quantities $F(\Gamma,\Gamma')$ have certain properties
that allow us to use the Hahn-Banach theorem to satisfy the inequalities
\eqref{(28)}, with constants $B(\Gamma)$ that depend linearly on $\Gamma$, in
the sense of \eqref{(23)}.  These properties, which follows 
immediately from the definition, are
\begin{eqnarray}F(\Gamma,\Gamma)&=&0 \label{Fa}\\
F(t\Gamma,t\Gamma')&=&tF(\Gamma,\Gamma')
\quad\quad {\rm for\ }t>0\label{Fb}
 \\
\label{Fc}F(\Gamma_1\times \Gamma_2,\Gamma_1'\times
\Gamma_2')& \leq& F(\Gamma_1,\Gamma_1')+F(\Gamma_2,\Gamma_2')
\\ \label{Fd}F(\Gamma\times
\Gamma_0,\Gamma'\times \Gamma_0)&= &F(\Gamma,\Gamma')\quad\quad
\hbox{\rm for all\ \ }\Gamma_0.
\end{eqnarray}
In fact, \eqref{Fa} and \eqref{Fb} are also shared by the $D$'s and
the $E$'s.  The `subadditivity' \eqref{Fc} holds also for the $E$'s,
but the `translational invariance' \eqref{Fd} might only hold for the
$F$'s.  Eq.\ \eqref{(29)}, and, more generally, the \lq triangle inequality\rq\
\begin{equation}\label{Fe}
F(\Gamma,\Gamma^{\prime\prime})\leq F(\Gamma,\Gamma')+
F(\Gamma',\Gamma^{\prime\prime})
\end{equation}
are simple consequences of \eqref{Fc} and 
\eqref{Fd}. 
Using these properties we can now derive

\begin{thm}{Theorem 7 {\bf (Universal entropy)}} The additive entropy
constants of all systems can be calibrated in such a way that the
entropy is additive and extensive, and $X\prec Y$ implies
$S(X)\leq S(Y)$, even when $X$ and $Y$ do not belong to the same state
space.
\end{thm}

\begin{proof} The proof is a simple application of the Hahn-Banach 
theorem.
Consider the set ${\mathcal S}$ of all pairs of state spaces 
$(\Gamma,\Gamma')$.
On ${\mathcal S}$ we define an equivalence relation by declaring
$(\Gamma,\Gamma')$ to be equivalent to $(\Gamma\times 
\Gamma_0,\Gamma'\times
\Gamma_0)$ for all $\Gamma_0$. Denote by $[\Gamma,\Gamma']$ the 
equivalence
class of $(\Gamma,\Gamma')$ and let ${\mathcal L}$ be the set of all these
equivalence classes.

On ${\mathcal L}$ we define  multiplication by scalars and addition in the 
following way:
\begin{eqnarray}
t[\Gamma,\Gamma']&:= &[t\Gamma,t\Gamma']\qquad\quad\hbox{\rm for }  
t> 0 
\\
t[\Gamma,\Gamma']&:= &[-t\Gamma',-t\Gamma]\qquad\hbox{\rm for } t< 0 
\\
0[\Gamma,\Gamma']&:= &[\Gamma,\Gamma]= [\Gamma',\Gamma'] \\
{\phantom +}[\Gamma_{1},\Gamma_{1}']
+[\Gamma_{2},\Gamma_{2}']&:=&
[\Gamma_{1}\times\Gamma_{2},\Gamma_{1}'\times\Gamma_{2}'].
\end{eqnarray}
With these operations ${\mathcal L}$ becomes a vector space, which is
infinite dimensional in general.  The zero element is the class
$[\Gamma,\Gamma]$ for any $\Gamma$, because by our definition of the
equivalence relation $(\Gamma,\Gamma)$ is equivalent to $(\Gamma\times
\Gamma',\Gamma\times \Gamma')$, which in turn is equivalent to
$(\Gamma',\Gamma')$.  Note that for the same reason $[\Gamma',\Gamma]$
is the negative of $[\Gamma,\Gamma']$.

Next, we define a function $H$ on ${\mathcal L}$ by
\begin{equation}H([\Gamma,\Gamma']):=F(\Gamma,\Gamma')\end{equation}
Because of \eqref{Fd}, this function is well defined and it takes values in 
$(-\infty,\infty]$. Moreover, it follows from \eqref{Fb} and \eqref{Fc} that $H$ is 
homogeneous, i.e., $H(t[\Gamma,\Gamma'])=tH([\Gamma,\Gamma'])$, 
and subadditive, i.e., $H([\Gamma_1,\Gamma_1']+[\Gamma_2,\Gamma_2'])
\leq H([\Gamma_1,\Gamma_1']) + H([\Gamma_2,\Gamma_2'])$. 
Likewise, 
\begin{equation}G([\Gamma,\Gamma']):=-F(\Gamma',\Gamma)\end{equation}
is homogeneous and superadditive, i.e., $G([\Gamma_1,\Gamma_1']+
[\Gamma_2,\Gamma_2']) \geq G([\Gamma_1,\Gamma_1']) 
+G([\Gamma_2,\Gamma_2'])$.
By \eqref{(29)}
we have $G\leq F$ so, by the Hahn-Banach theorem, there exists a 
real-valued 
{\it linear} function $L$ on ${\mathcal L}$ lying between $G$ and $H$; that is
\begin{equation}\label{between}
-F(\Gamma',\Gamma) \leq L([\Gamma,\Gamma']) \leq F(\Gamma,\Gamma').
\end{equation}

Pick any 
fixed $\Gamma_0$ and define 
\begin{equation}B(\Gamma):=L([\Gamma_0\times\Gamma,\Gamma_0]).
\end{equation}
By linearity, $L$ satisfies $L([\Gamma,\Gamma']) = -L(-[\Gamma,\Gamma'])
=-L([\Gamma',\Gamma])$. 
We then have 
\begin{equation}B(\Gamma)-B(\Gamma')=L([\Gamma_0
    \times\Gamma,\Gamma_0])+
L([\Gamma_0, \Gamma_{0}\times \Gamma'])=L([\Gamma,\Gamma'])
\end{equation}
and hence \eqref{(28)} is satisfied.
\end{proof}

Our final remark concerns the remaining non-uniqueness of the constants 
$B(\Gamma)$.
This indeterminacy  can be traced back to the non-uniqueness of
a linear functional lying between $-F(\Gamma',\Gamma)$ and
$F(\Gamma,\Gamma')$ and has two
possible sources: One is that some pairs of state-spaces $\Gamma$ and
$\Gamma'$ may not be connected, i.e., $F(\Gamma,\Gamma')$ may be
infinite (in which case $F(\Gamma',\Gamma)$ is also infinite by axiom
A16). 
The other is that there might be a true gap, i.e.,
\begin{equation}\label{(32)}
-F(\Gamma',\Gamma)~<~F(\Gamma,\Gamma')
\end{equation}
might hold for some state spaces, even if both sides are finite.
 
In nature only states containing the same amount of the chemical
elements can be transformed into each other. Hence
$F(\Gamma,\Gamma')=+\infty$ for many pairs of state spaces, in
particular, for those that contain different amounts of some chemical
element.  The constants $B(\Gamma)$ are, therefore, never unique: For
each equivalence class of state spaces (with respect to the relation of
connectedness) one can define a constant that is arbitrary except for
the proviso that the constants should be additive and extensive under
composition and scaling of systems. In our world there are  92 chemical
elements (or, strictly speaking, a somewhat larger number, $N$, since
one should count different isotopes as different elements), and this
leaves us with at least 92 free constants that specify the entropy of
one gram of each of the chemical elements in some specific state.  

The other possible source of non-uniqueness, a nontrivial gap
\eqref{(32)} for systems with the same composition in terms of the
chemical elements is, as far as we know, not realized in nature,
although it is a logical possibility.  The true situation seems rather
to be the following: Every state space $\Gamma$ is connected to a
distinguished state space
\begin{equation}
\Lambda(\Gamma)=\lambda_1\Gamma_1\times\dots\times\lambda_N\Gamma_N
\end{equation}
where the $\Gamma_i$ are the state spaces of one mole of each of the 
chemical
elements, and the numbers $(\lambda_1,\dots,\lambda_N)$ specify the 
amount 
of
each chemical element in $\Gamma$. We have 
\begin{equation}\label{la}\Lambda(t\Gamma)=t\Lambda(\Gamma)
\end{equation}
and
\begin{equation}\label{lb}\Lambda(\Gamma\times\Gamma')=\Lambda(\Gamma) 
\times\Lambda(\Gamma').
\end{equation}
Moreover (and this is the crucial `experimental fact'), 
\begin{equation}\label{lc}-F(\Lambda(\Gamma),\Gamma)=F(\Gamma, \Lambda(\Gamma))
\end{equation}
for all $\Gamma$. Note that \eqref{lc} is subject to experimental
verification by measuring on the one hand entropy differences for
processes that synthesize chemical compounds from the elements
(possibly through many intermediate steps and with the aid of
catalysts), and on the other hand for processes where chemical
compounds are decomposed into the elements. This procedure need not
invoke any semipermeable membranes.

It follows from \eqref{(29)} \eqref{Fe} and \eqref{lc}
that 
\begin{equation}\label{lddd}F(\Gamma,\Gamma')=F(\Gamma,\Lambda(\Gamma))+
F(\Lambda(\Gamma),
\Gamma')\end{equation}
and
\begin{equation}\label{ld}-F(\Gamma',\Gamma)=F(\Gamma,\Gamma')
\end{equation}
for all $\Gamma'$ that are connected to $\Gamma$.
Moreover, an explicit formula for $B(\Gamma)$ can be given:
\begin{equation}\label{le}B(\Gamma)=F(\Gamma, \Lambda(\Gamma).
\end{equation}
If $F(\Gamma,\Gamma')=\infty$, then \eqref{(28)} holds trivially, while 
for connected state spaces
$\Gamma$ and $\Gamma'$ we have by \eqref{lddd} and \eqref{ld}
\begin{equation}\label{lee}B(\Gamma)-B(\Gamma')=
F(\Gamma,\Gamma')=-F(\Gamma',\Gamma),\end{equation}
i.e., the inequalities \eqref{(28)} are saturated. It is also clear that 
in this case 
$B(\Gamma)$ is unique up to the choice of arbitrary constants for the 
fixed 
systems $\Gamma_1,\dots,\Gamma_N$. The particular choice \eqref{le} 
corresponds 
to 
putting $B(\Gamma_i)=0$ for the chemical elements $i=1,\dots,N$.


In conclusion, once the 
entropy constants for the chemical elements have been fixed and a
temperature unit has been chosen (to fix the 
multiplicative constants) the universal entropy is completely fixed.

\smallskip
\noindent  {\bf Acknowledgements.} We are indebted to many people
for helpful discussions, including Fred Almgren, Thor Bak, Bernard
Baumgartner, Pierluigi Contucci, Roy Jackson, Anthony Knapp, Martin
Kruskal, Mary Beth Ruskai and Jan Philip Solovej. We thank Daniel Goroff
for drawing our attention to reference \cite{HM}.

\section{Some Speculations and Open Problems} \label{sec2}
1.\quad
As we have stressed, the purpose of the entropy function is to
quantify the list of equilibrium states that can evolve from other
equilibrium states. The evolution can be arbitrarily violent, but always
$S(X) \leq S(Y)$ if $X\prec Y$. Indeed, the early thermodynamicists
understood the meaning of entropy as defined for equilibrium states. Of
course, in the real world, one is often close to equilibrium without
actually being there, and it makes sense to talk about entropy as a
function of time, and even space, for situations close to equilibrium.
We do a similar thing with respect to temperature, which
has the same problem that temperature is only strictly
defined for a homogeneous system in equilibrium.
At some point the thought arose (and we confess
our ignorance about how it arose and by whom) that it ought to be
possible to define an entropy function rigorously for manifestly {\em
non-}equilibrium states in such a way that the numerical value of this
function will increase with time as a system goes from one equilibrium
state to another.

Despite the fact that most physicists believe in such a non-equilibrium
entropy it has so far proved to be impossible to define it in a clearly
satisfactory  way.  (For example  Boltzmann's famous
H-Theorem shows the steady increase of a certain function called $H$.
This, however, is not the whole story, as Boltzmann
himself knew; for one thing, $H \neq S$ in equilibrium
(except for ideal gases), and, for another, no one has so
far proved the increase without making severe
assumptions, and then only for a short time interval
(cf.\ \cite{[11]}).)
Even today, there is no
universal agreement about what, precisely,  one should try to prove (as
an example of the ongoing discussion, see \cite{[12]}).

It is not clear if entropy can be consistently extended to
non-equilibrium situations in the desired way. After a century and a
half of thought, including the rise of the science of statistical
mechanics as a paradigm (which was not available to the early
thermodynamicists and, therefore, outside their thoughts),  we are far
from success. It has to be added, however, that a great deal of
progress in understanding the problem has been made recently (e.g.,
\cite{[13]}).

If such a concept can be formulated precisely, will it have to involve
the notion of atoms and statistical mechanical concepts, or can it be
defined irrespective of models, as we have done for the entropy of
equilibrium states? This is the question we pose. 

There are several major problems to be overcome, and we list two of
them.
\begin{enumerate}
\item[a.] The problem of time reversibility: If the concept is going to 
depend upon mechanical models, we have  to account for the fact that 
both classical and quantum mechanics are time reversible. This makes
it difficult to construct a mechanical quantity that can only increase 
under classical or quantum mechanical time evolution. Indeed, this
problem usually occupies center stage in most discussions of the
subject, but it might, ultimately, not be the most difficult problem
after all.
\item[b.] In our view of the subject, a key role is played by the idea
of a more or less arbitrary (peaceful or violent) interaction of the
system under discussion with the rest of the universe whose final
result is a change of the system, the change in height of a weight, and
nothing more.  How can one model such an arbitrary interaction in a
mechanical way? By means of an additional term in the Hamiltonian? That
hardly seems like a reasonable way to model a sledgehammer that happens
to fall on the system or a gorilla jumping up and down.  
Most discussions of entropy increase refer to the evolution of 
systems subject to  a  dynamical evolution that is usually
Hamiltonian (possibly time dependent) or a mixture of Hamiltonian
and stochastic evolution.  This
can hardly even cope with describing a steam engine, much less a
random, violent external force.
\end{enumerate}

As a matter of fact, most people would recognize a) as the important
problem, now and in the past. In b) we interject a new note, which, to
us, is possibly more difficult. There are several proposals for a
resolution of the irreversibility problem, such as the large number 
($10^{23} \approx \infty$) of atoms involved, or the `sensitive
dependence on initial conditions' (one can shoot a ball out of a
cannon, but it is very difficult to shoot it back into the cannon's
mouth). Problem b), in contrast, has not received nearly as much attention.
\medskip 

2.\quad
An essential role in  our story was played by axioms A4
and A5, which require the possiblity of having
arbitrarily small samples of a given material and that
these small samples behave in exactly the same way as a 1
kilogram sample.
While this assumption is
made in everyone's formulation of the second law, we have to recognize
that absurdities will arise if we push the concept to its extreme. 
Eventually the atomic nature of matter will reveal itself and entropy
will cease to have a clear meaning. What protects us is the huge power
of ten (e.g., $10^{23}$) that separates macroscopic physics and the
realm of atoms.  

Likewise, a huge power of ten separates time scales that make physical
sense in the ordinary macroscopic world and time scales (such as
$10^{25}$ seconds $= 10^7$ times the age of the universe) which are
needed for atomic fluctuations to upset the time evolution one would
obtain from macroscopic dynamics. One might say that one of the hidden
assumptions in our (and everyone else's) analysis is that {\em $\prec$
is reproducible}, i.e., $X\prec Y$ either holds or it does not, and
there are no hidden stochastic or probabilistic mechanisms that would
make the list of pairs $X\prec Y$ `fuzzy'.

One of the burgeoning area of physics research is 'mesoscopics', which
deals with the interesting properties of tiny pieces of matter that
might contain only a million atoms (= a cube of 100 atoms on a side) or
less. At some point the second law has to get fuzzy and a significant
open problem is to formulate a fuzzy version of what we have done in
\cite{[7]}.
Of course, no amount of ingenuity with mesoscopic systems is
allowed to violate the second law on the macroscopic level, and this
will have to be taken into account. One possibility could  be that
an entropy function can still be defined for mesoscopic systems but that
$\prec$ is fuzzy, with the consequence that entropy increases only on
`the average', but in a totally unpredictable way -- so that the
occasional decrease of entropy cannot be utilized to violate the second
law on the macroscopic level. 

There are other problems as well. A simple system, such as a container
of hydrogen gas, has states described by energy and volume. For a
mesoscopic quantity of matter, this may not suffice to describe an
equilibrium state.  Another problem is the meaning of equilibrium and
the implicit assumption we made that after the (violent) adiabatic
process is over  the system will eventually come to some equilibrium
state in a time scale that is short compared to the age of the
universe. On the mesoscopic level, the achievement of equilibrium may
be more delicate  because a mesoscopic system might never settle down
to a state with insignificant fluctuations that one would be pleased to
call an equilibrium state.

\bigskip
To summarize, we have listed two (and there are surely more) areas
in which more thought, both mathematical and physical, is needed:
the extension of the second law and the entropy concept to 
1) non-equilibrium situations and 2) mesoscopic and even atomic
situations. One might object that the problems cannot be solved 
until the `rules of the game' are made clear, but discovering the 
rules is part of the problem. That is sometimes inherent in
mathematical physics, and that is one of the intellectual challenges 
of the field.

\section{Some Remarks About Statistical Mechanics} \label{sec3}

We are frequently asked about the connection between our
approach to the second law of thermodynamics and the statistical
mechanical Boltzmann-Gibbs-Maxwell approach. Let us make it clear that
we  value statistical mechanics as much as any physicist.  It  is a
powerful tool for understanding physical phenomena and for calculating
many quantities, especially in systems at or near equilibrium. It is used
to calculate entropy, specific and latent heats, phase transition
properties, transport coefficients and so on, often with good accuracy.
Important examples abound, such as Max Planck's 1901 realization 
\cite{furnace} that by
staring into a furnace he could find Avogadro's number, or Linus Pauling's
highly accurate back-of-the-envelope calculation of the residual entropy
of ice \cite{ice} in 1935.  
But is statistical mechanics essential for the second law?

In any event, it is still beyond anyone's computational ability (except
in idealized situations) to account for this very precise, essentially
infinitely accurate law of physics from statistical mechanical principles.
No exception has ever been found to the second law of thermodynamics---not
even a tiny one.  Like conservation of energy (the ``first'' law) the
existence of a law so precise and so independent of details of models
must have a logical foundation that is independent of the fact that
matter is composed of interacting particles.  Our aim in \cite{[7]}
was to explore that foundation. It was also our aim to try to formulate
clear statements on the macroscopic level so that statistical mechanics
can try to explain them in microscopic terms.

As Albert Einstein put it \cite{E}, ``A theory is the more impressive the
greater the simplicity of its premises is, the more different kinds of
things it relates, and the more extended is its area of applicability.
Therefore the deep impression which classical thermodynamics made upon me.
It is the only physical theory of universal content concerning which
I am convinced that, within the framework of the applicability of its
basic concepts, it will never be overthrown''.

We maintain, that the second law, as understood for equilibrium states
of macroscopic systems, does not require statistical mechanics, or any
other particular mechanics, for its existence. It does require certain
properties of macroscopic systems, and statistical mechanics is one
model that, hopefully, can give those properties, such as
irreversibility. One should not confuse the existence, importance, and
usefulness of the Boltzmann-Gibbs-Maxwell theory with its necessity on
the macroscopic level as far as the second law is concerned. Another
way to make the point is this: If the statistical mechanics of atoms
is essential for the second law, then that law must imply something
about atoms and their dynamics. Does the second law prove the
existence of atoms in the way that light scattering, for example,
tells us what Avogadro's number has to be? Does the law distinguish
between classical and quantum mechanics? The answer to these and
similar questions is \lq\lq no\rq\rq\ and, if there were a direct
connection, the late 19th-century wars about the existence of atoms
would have been won much sooner. Alas, there is no such direct
connection that we are aware of, despite the many examples in which
atomic constants make an appearance at the macroscopic level such as
Planck's radiation formula mentioned above, the Sackur-Tetrode
equation, stability of matter with Coulomb forces, and so on. The
second law, however, is not such an example.

\bigskip\medskip

\noindent
{\sc Elliot H.\ Lieb},
Depts. of Mathematics and Physics,
 Princeton University,
Jadwin Hall,  P.O. Box 708,
Princeton, NJ  08544,
USA
\medskip

\noindent
{\sc Jakob Yngvason},
Institut f\"ur Theoretische Physik,
 Universit\"at Wien,
Boltzmanngasse 5,
 A 1090 Vienna,
 Austria
\end{document}